\documentclass[aps,prb,preprint]{revtex4-2}
\usepackage{graphicx} 

\usepackage{ulem}
\usepackage{amssymb}
\usepackage{xcolor}
\usepackage{amsmath}
\usepackage{bm}
\usepackage{graphicx}
\usepackage{bm}
\usepackage{epsfig}
\usepackage{epstopdf}
\usepackage{siunitx}
\usepackage{color}
\usepackage{float}

\usepackage[top=1cm, bottom=2cm, left=3cm, right=1cm]{geometry}

\begin{document}

\title{Band offsets in InP/ZnSe nanocrystals evaluated using two-photon transitions analysis.}
\author{K.I. Russkikh, A.A. Golovatenko, A.V. Rodina }
\affiliation{Ioffe Institute, Russian Academy of Sciences, 194021 St.~Petersburg, Russia}

\begin{abstract}
We present a semi-analytical theoretical $kp$-study of the energy structure and optical transitions in spherical core-shell InP/ZnSe nanocrystals. We use the eight-band Kane model and the six-band Luttinger Hamiltonian in the spherical approximation to calculate the electron and hole energy spectra, respectively. The influence of the Coulomb interaction is considered perturbatively.
The one- and two-photon absorption spectra are calculated as functions of the band offsets between the InP core and ZnSe shell. Exciton states responsible for the main features in the two-photon absorption spectra of InP/ZnSe nanocrystals are identified and the spectral dependence of the linear-circular dichroism signal is predicted. We show that in the presence of inhomogeneous broadening, the transition to the ground two-photon-active exciton state can be hidden behind intense transitions to higher-lying states.  A comparison of the calculated one- and two-photon absorption spectra with the available experimental data shows that, depending on the lattice strain in the InP core, the range of possible valence band offsets is $0.85-1$~eV.   The determined range exceeds the natural valence band offset of \(0.57\)~eV expected from first-principles calculations, indicating the presence of electric dipoles formed by preferential Zn-P bonds at the InP/ZnSe heterointerface.

\end{abstract}





\maketitle

\section{Introduction}
Today, the importance of research into colloidal quantum dots, also known as nanocrystals (NCs), requires no further justification following the awarding of the 2023 Nobel Prize in Chemistry for their discovery. Colloidal synthesis allows for the formation of NCs from different materials (IV, II-VI, III-V semiconductors, lead-halide perovskites) with a wide range of sizes. These two factors allow for tuning of optical absorption and emission spectra, and together with a high quantum efficiency, make NCs attractive for numerous applications. Within this family of colloidal NCs, InP-based NCs promote tunable photoluminescence (PL) from the visible to the near-infrared spectrum. 

In this family of colloidal nanocrystals, InP-based NCs possess high photoluminescence quantum yield and  narrow emission bandwidths,\cite{Efros2021,Jang2020,Xu2018,Kim2024} making them highly suitable for next-generation optoelectronic applications, including high-resolution displays \cite{Won2019, Efros2019}, new types of lasers\cite{Klimov2021, Klimov2023, Chong2026}, wearable electronics, and augmented and virtual reality technologies \cite{Park2023,Liu2022,Yang2020}. The low toxicity of InP nanocrystals allows for their application in bacterial monitoring \cite{Yousefi2024}, and  their solution-processability facilitates their application in cost-effective, large-scale production, particularly appealing for flexible and stretchable electronics\cite{Fan2024,Hua2024,Lee2023}. 

\begin{figure*}[h!]
 \centering \includegraphics[width=0.8\linewidth]{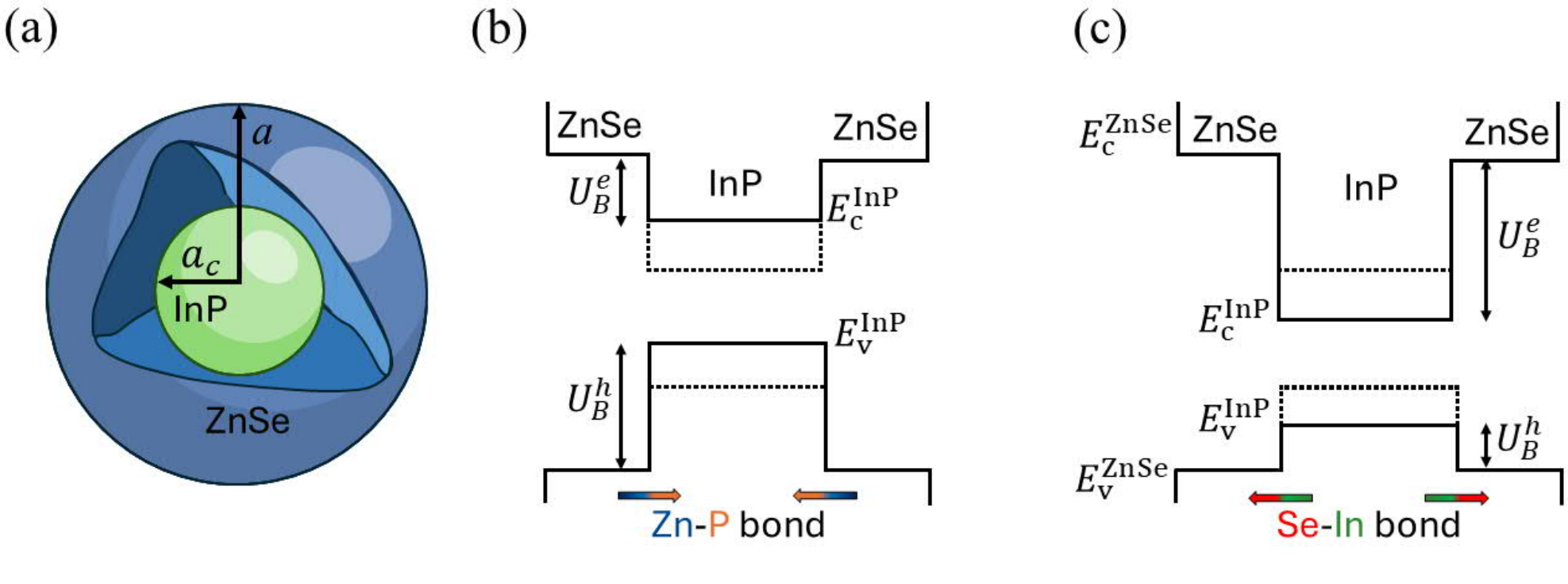}
 \caption{ (a) Schematic of an InP/ZnSe NC with core radius $a_c$ and the total NC radius $a$. Energy profiles  for electrons with barrier (band offset)  $U^e_B=E_c^{\rm ZnSe} -E_c^{\rm InP}$ and hole with barrier $U_B^h = E_v^{\rm InP} -E_v^{\rm ZnSe}$ corresponding to the case of preferential P-Zn (b) or In-Se (c) bonds at the InP/ZnSe heterointerface, respectively. Gradient arrows show the direction of the sum interface dipole. The dotted lines show the natural band offsets defined as the differences of the bulk energy band extrema with respect to the vacuum level.}
 \label{fig:Fig1}
\end{figure*}

Despite numerous advances in the synthesis of InP nanocrystals, there is still room for improvement. InP nanocrystals have a rich history of theoretical \cite{Richard1996,Efros1998, Fu1998,Franceschetti1999} and experimental \cite{Micic1994,Wells1995,Bertram1998} research dating back to the end of the last millennium. Since bare-core InP nanocrystals exhibit low photoluminescence quantum yield, significant efforts have been made to improve it by growing a ZnSe or ZnS shell. The successful implementation of ZnSe shell growth in Ref.~\cite{Won2019} allowed for an increase in the photoluminescence quantum yield and provided new impetus to research on InP-based nanocrystals \cite{Shin2024, Soheyli2023, Yu2021, Baek2024, Van2022, Li2022, Brodu2019,Rafipoor2019,Velosa2022,Tolmachev2024,Respekta2025,Schiettecatte2024,Giordano2025,Schiettecatte2025}. 

Recent articles on the analysis of InP/ZnSe and InP/ZnSe/ZnS NCs by different optical methods, such as the Stokes shift of the non-resonant photoluminescence \cite{Cavanaugh2023}, one-photon photoluminescence excitation (1PLE) and two-photon photoluminescence excitation (2PLE) \cite{Tolmachev2024,Respekta2025}, transient differential absorption \cite{Velosa2022, Respekta2025} and Raman spectroscopy \cite{Rafipoor2019} raised the issue of the conduction and valence band offsets between InP and ZnSe. 

The uncertainty in the band offset values in InP/ZnSe NCs has two origins. First, due to the absence of a common atom in the InP/ZnSe pair, an electric dipole field arises at the  heterointerface between the InP core and ZnSe shell \cite{Jeong2022}.  
An example of a spherical core-shell InP/ZnSe NC and energy profiles for the electron and hole modified by the presence of electric dipoles at the heterointerface are shown in Fig.~\ref{fig:Fig1}.
The direction of the dipole field depends on the type of bonds at the heterointerface \cite{Jeong2022}. The sum dipole is directed from the shell to the core in the case of preferential P-Zn bonds (see Fig.~\ref{fig:Fig1}(b)) and from the core to the shell in the case of preferential In-Se bonds (see Fig.~\ref{fig:Fig1}(c)) at the heterointerface. According to Ref.~\cite{Jeong2022}, the formation of dipoles at the heterointerface results in a shift of the conduction and valence bands of the InP core with respect to the vacuum level. Dipoles formed by P-Zn bonds increase the valence band offset between InP and ZnSe, while dipoles formed by In-Se bonds decrease the valence band offset, and vise versa for the conduction band offset.  The resulting band offsets for the electron and hole are different from the natural band offsets\cite{Wei1998} (see dotted lines in Figs.~\ref{fig:Fig1}(b,c)). This effect is similar to the effect of surface dipoles formed by polar ligands in bare-core nanocrystals \cite{Boles2016,Morgan2018}.

The second source of uncertainty in the band offsets is the compressive strain in the InP core and the tensile strain in the ZnSe shell, caused by the $3.5\%$ lattice mismatch \cite{Lange2020}. The strain changes the interatomic distance and leads to shifts in the conduction and valence band energies.  In Ref.~\cite{Lange2020}, the conduction band offset of $0.41$~eV was determined based on modeling the dependence of the Raman shift on the ZnSe shell thickness \cite{Rafipoor2019} in InP/ZnSe NCs.   This value is consistent with the data from Ref.~\cite{Stevanovic2014} and differs from the natural band offset between InP and ZnSe .~\cite{Wei1998}

Recently, the valence band offset values in InP/ZnSe NCs were estimated based on the theoretical analysis of 1PLE, 2PLE, and transient differential absorption spectra in the framework of numerical $kp$-calculations. \cite{Respekta2025} Calculation of the one-photon transition energies in Ref.~\cite{Respekta2025} was performed numerically using a one-band model for electrons and the six-band Luttinger Hamiltonian for holes, without accounting for the electron-hole Coulomb interaction. The determined lower limit for the valence band offset of $0.693$~eV \cite{Respekta2025} is consistent with the conclusion drawn from the modeling of the Raman shifts in Ref.~\cite{Lange2020}, but contradicts the conclusions about a low valence band offset made in  Ref.~\cite{Velosa2022}. 
Calculations from \cite{Respekta2025} explained the experimental results for the 1PLE  and transient differential absorption data. However, the 2PLE spectrum of InP/ZnSe NCs shown in Ref.~\cite{Respekta2025} remained unexplained, including the absorption feature at $\approx 2.4$~eV, which was also observed in the $2$PLE spectrum of InP/ZnSe/ZnS with a similar InP core radius in Ref.~\cite{Tolmachev2024}.    

In contrast to one-photon spectroscopy, two-photon spectroscopy provides information about transitions between levels with opposite spatial parity. Together, these two methods make it possible to trace the entire exciton energy spectrum, as was done previously for CdSe \cite{Blanton1996},chalcopyrite\cite{Nagamine2018} and lead-halide perovskite NCs \cite{Bataev2026}. Theoretical calculations of two-photon optical transitions are more elaborate compared to one-photon transitions, especially when the wave functions of charge carriers are calculated numerically. 

In this paper, we show that analytical $kp$-analysis allows for the calculation of the exciton energy spectrum, as well as one- and two-photon transition probabilities in InP/ZnSe NCs. We consider a model of spherical NCs and apply a conventional approach based on the $kp$-method \cite{Ekimov1993,Efros1998,Sercel1999,Pokatilov2001P245328}
for the calculation of the electron and hole energy spectrum, as well as wave functions for variable valence and conduction band offsets. Using these wave functions, we calculate the Coulomb corrections to the exciton energies and the probabilities of one- and two-photon optical transitions. The calculated one- and two-photon absorption spectra for different values of the valence band offset are compared with the experimental data from Ref.~\cite{Respekta2025}. The effect of linear-circular dichroism, previously observed in bulk semiconductors \cite{Ivchenko1973,Dvornikov1978,Mozol1980,Ganichev1993} and predicted theoretically for perovskite NCs \cite{Blundell2021} is studied. We show that joint modeling of 1PLE and 2PLE data imposes strict constraints on the range of suitable band offset values. Finally, we estimate contributions of the dipoles at the InP/ZnSe heterointerface and the lattice strain to the difference between the determined and natural band offsets.
\section{Theoretical model}
We consider the InP/ZnSe NC as two concentric spheres, with $a_c$ being the radius of the InP core and $a$ being the total NC radius (see Fig.~\ref{fig:Fig1}(a)). Although the actual shape of InP/ZnSe NCs is close to tetrahedral \cite{Won2019, Respekta2025},
recent calculations \cite{Respekta2025,Planelles2026} show that the model of a spherical NC is a good approximation for estimating the energy structure and optical properties of exciton states. Equation \eqref{eq1} defines spherically symmetric potential energy profiles for electrons, $U_{e}(r)$, and holes , $U_{h}(r)$, shown in Fig.~\ref{fig:Fig1}(b,c):
\begin{eqnarray}\label{eq1}
U_{e,h}(r)=\begin{cases}0 & r < a_{c}\\U_{B}^{e,h} & a_c\leq r\leq a\\+\infty & r > a\end{cases}.
\end{eqnarray}
The barriers between the InP core and the ZnSe shell for electrons, $U^e_B=E_c^{\rm ZnSe} -E_c^{\rm InP}$, and holes, $U_B^h = E_v^{\rm InP} -E_v^{\rm ZnSe}$, are considered as variable parameters due to the unknown contributions of dipoles at the heterointerface and lattice strain. The bottom of the conduction band, $E_c^{\rm ZnSe}$, and the top of the valence band, $E_v^{\rm ZnSe}$, in the ZnSe shell are assumed to be the same as in  bulk. 
The bottom of the conduction band, $E_c^{\rm InP}$, and the top of the valence band, $E_v^{\rm InP}$, in the InP core can differ from the bulk values $E_{\text{c},\text{bulk}}^{\text{InP}}$ and $E_{\text{v},\text{bulk}}^{\text{InP}}$ due to the electric dipoles and lattice strain effects.  We consider both barriers $U_B^{e,h}\geq0$, i.e. the NC energy structure is either type-I or quasi-type-II. The total sum of conduction and valence band offsets is assumed to be constant $U_B^h+U_B^e=E_g^{\text{ZnSe}}-E_g^{\text{InP}}$, where $E_g^{\text{InP}}$ and $E_g^{\text{ZnSe}}$ are the band-gaps of InP and ZnSe, respectively. This assumption is justified in the case of the influence of the interface electric dipoles, while lattice strain can additionally modify the  band gap energies of InP and ZnSe. We neglect this effect in our calculations and assume that the band gaps remain constant. Infinite potential barriers for both electrons and holes are considered at the outer boundary of the ZnSe shell.

To calculate the electron energy $E_{\text{e}}$, which is measured from the bottom of the InP conduction band  $E_{\text{c}}^{\text{InP}}$, we use the eight-band $\bm{k}\cdot\bm{p}$ Kane model  \cite{Rodina2003} (see ESI section S1 for details). Within this model, the total electron eight-band component wave function $\bm{\Psi}_e=\{\Psi^c_e,\bm{\Psi}^v_e\}$ consists of a two-band (scalar spinor) $\Psi^c_e$ conduction band and a six-band (vector spinor) $\bm{\Psi}^v_e$ valence band components. They are written  in the basis of the Bloch functions of the conduction band, $|S\rangle u_{\pm1/2}$, and the valence band, $\{|X\rangle u_{\pm1/2},\ |Y\rangle u_{\pm1/2},\ |Z\rangle u_{\pm1/2}\}$, respectively.  Here $u_{\pm1/2}$ are the eigenvectors of the spin  operator $\hat{\bm S}$ (for $S=1/2$), and the conduction, $|S\rangle$, and  the valence band, $\{|X\rangle,\ |Y\rangle,\ |Z\rangle\}$, the orbital Bloch functions at the $\Gamma$ point of the Brillouin zone described by the internal angular momentum $\bm{I}$  for  $I=0$ and $I=1$, respectively  \cite{Rodina2003,Ivchenko2005}. The main contribution to $\bm{\Psi}_e$ comes from the conduction band spinor $\Psi^c_e$, which satisfies a Shr{\"o}dinger-like equation with the  energy-dependent  electron effective-mass $m_c(E_e,U_e)$ and the potential energy $U_e(r)$  (see ESI section S1 for details). 
As we discuss in Section \ref{results}, accounting for nonparabolic electron dispersion within the Kane model leads to lower energies of the electron excited states compared to the effective mass model with parabolic energy dispersion.

To calculate the hole energy spectrum, we use the six-band Luttinger Hamiltonian $\hat{H}_{6\times6}$ \cite{Luttinger1956} in the spherical approximation, i.e. neglecting the cubic symmetry of the hole energy dispersion  (see ESI section S2 for details). It has been shown \cite{Balderschi1974} that if we take $\gamma=(2\gamma_2+3\gamma_3)/5$, where $\gamma_2$, $\gamma_3$ are Luttinger parameters \cite{Luttinger1956} (see ESI section S2 for details), then the energy levels in any spherically symmetric confinement potential are correct up to the first order of perturbation theory. The second order correction is usually small because $\gamma_3-\gamma_2\ll\gamma_2+\gamma_3$. The hole envelope six-band component (vector spinor) wavefunction $\bm{\Psi}_h$ is considered in the hole representation, and the positive hole energy $E_{\text{h}}$ is measured from the top of the InP valence band $E_{\text{v}}^{\text{InP}}$ . 

The boundary conditions at the spherical InP/ZnSe heterointerface require the conservation of the flux density projection onto the unit normal vector to the interface $\bm{\tau}={\bm r}/|\bm{r}|$ (see ESI section S3 for details). We use standard boundary conditions that conserve the value of the electron envelope function $\Psi^c_e$ and the normal velocity component $V^e_{\bm{\tau}}$:

\begin{equation}\label{Eq2}
\Psi^c_e =\text{const}, \quad  V_{\bm{\tau}}^e \equiv (\bm{\tau}\cdot \hat{\bm{V}}_e){\bm \Psi}_e= \frac{\hbar}{m_c(E_\text{e},U_e)}(\bm{\tau}\cdot \hat {\bm k}\Psi^c_e)=\text{const}.
\end{equation}
Here $\hat {\bm k} = -i \bm{\nabla}$ is the wave vector operator, and in $V_{\bm{\tau}}^e$ we neglect the spin-dependent term, which, as shown in Ref. \cite{Rodina2003}, can be considered perturbatively.
 
The standard boundary conditions for the hole conserve each component of the vector $\bm{\Psi}_h$ and each component of its normal velocity $\bm{V}_{\tau}^h$ and are given by:
\begin{equation}\label{Eq3}
\bm{\Psi}_h =\text{const}, \quad  \bm{V}_{\bm{\tau}}^h \equiv (\bm{\tau}\cdot \hat{\bm{V}}_h){\bm \Psi}_h= (\bm{\tau}\cdot\frac{\partial \hat{H}_{6\times6}}{\partial {\bm k}}){\bm \Psi}_h=\text{const}.
\end{equation}
 Since we impose infinite potential barriers at the outer boundary of the ZnSe shell, the following conditions must be satisfied: $\Psi^c_e=0$ and $\bm{\Psi}_h=0$. 
Due to the spherical symmetry of the structure, the electron and hole states are  characterized by 
the total angular momentum $\bm{j}=\bm{J}+\bm{L}$ and its projection $m=j_z$, where $\bm{J}=\bm{I}+\bm{S}$ is the total internal angular momentum   and  $\bm{L}$ is the orbital angular momentum of the envelope wavefunction (see ESI section S4 for details) . This allows us to express the electron and hole wavefunctions as products of the angular and radial components (see ESI section S5 for details).  Substituting these wavefunctions into the boundary conditions \eqref{Eq2},\eqref{Eq3}, we obtain analytical systems of equations for the calculation of the quantum confinement energies $E_{\text{e}}$ and $E_{\text{h}}$. 

The energy levels of the electron and hole are denoted as  $n_eQ_e$  and $n_h\tilde Q_{j_h}$ respectively, where $n_{e,h}$ are the principal quantum numbers. $Q$ and $\tilde Q$ denote the symmetries corresponding to the lowest angular momentum quantum number $l$ of the  electron and hole envelope  wave functions ($Q, \tilde Q=S,P,D$
for $l=0,1,2$, respectively), and $j_h$  is the total angular momentum of the hole. Following Ref. \cite{Ekimov1993}, we use the notation $n_eQ_e$ for electrons without specifying the angular momentum $j_e$, since  the spin-orbit splitting of the electron states $nQ_{e}$ with $l>0$ by the value of total angular momentum $j_e=l\pm1/2$ is small. This splitting is induced by the spherical boundary \cite{Ekimov1993, Pokatilov2001P245328} and can be treated in the eight-band Kane model as a perturbation \cite{Rodina2003}. We note that in Ref. \cite{Richard1996}, an alternative notation for hole levels was used, which displays all the angular harmonics included in the hole wavefunction.

The total exciton optical transition energy $E_\text{{X}}\equiv E_{\text{X}}(n_eQ_en_h\tilde Q_{j_h})$ is given by:
\begin{eqnarray}\label{eqEx}
E_{\text{X}} =E_{\text{e}}+E_{\text{h}}+E_g^{\text{InP}}+E_{\text{Coul}},
\end{eqnarray}
where $E_\text{e}\equiv E_\text{e}(n_eQ_e)$ and $E_{\text{h}}\equiv E_{\text{h}}(n_h\tilde Q_{j_h})$ are the electron and hole energies, respectively, and $E_{\text{Coul}} \equiv E_{\text{Coul}}(n_eQ_e,n_h\tilde Q_{j_h})$ is the energy of the Coulomb interaction between the electron and hole. We consider the strong spatial confinement regime when the Coulomb energy correction to the exciton energy can be treated as a perturbation $E_{\text{Coul}}=\langle\bm{\Psi}_{\text{ex}}(\bm{r}_e,\bm{r}_h)|\hat{V}_{\text{Coul}}|\bm{\Psi}_{\text{ex}}(\bm{r}_e,\bm{r}_h)\rangle$, where $\hat{V}_{\text{Coul}}=-e^2/[
\varepsilon_{\text{av}}|\bm{r}_e-\bm{r}_h|]$ and $\varepsilon_{\text{av}}=8.3$ is the average high frequency dielectric constant  (see ESI section S6 for details), and the  exciton wavefunction is   factorized as $\bm{\Psi}_{\text{ex}}(\bm{r}_e,\bm{r}_h)=\Psi^c_e(\bm{r}_e) \bm{\Psi}_{h}(\bm{r}_h)$. The treatment of the Coulomb interaction in the strong confinement regime as a perturbation is consistent with the conclusions of recent numerical calculations for InP/ZnSe NCs within the interaction configuration framework from Ref.~\cite{Planelles2026}.

We consider the  probabilities of one- and two-photon absorption in the dipole approximation. For one-photon transitions, we calculate matrix elements between electron and hole eigen-states for the following operator: 
$\hat{\mathcal{V}}_{\alpha}^{cv}=-e\hat{\bm{p}}\cdot \bm{A}_{\alpha}/cm_0$, where $\bm{A}_{\alpha}=A_0\bm{e}_{\alpha}$  is the electromagnetic field vector potential with amplitude $A_0$ and circular, $\bm{e}_\pm=1/\sqrt{2}\{1,\pm i,0\},$
or linear, $\bm{e}_x=\{1,0,0\}, \ \bm{e}_y=\{0,1,0\}$, polarization vector for the light propagating along the $z$ direction. For the spherical NCs considered in this paper, the $z'$ axis of the NC frame can be chosen to coincide with the $z$ axis of the laboratory frame. In the presence of NC shape anisotropy (oblate or prolate), the $z$ and $z'$ axes can have arbitrary relative orientations, which leads to the appearance of an additional nonzero matrix element for the $\bm {e}_z$ light polarization. Also, the averaging over the all possible NC spatial orientations is required in this case.
The  momentum operator $\hat{\bm{p}}=-i\hbar\bm{\nabla}$ acts on the Bloch components of the electron and hole wavefunctions. 

To model the differential absorption spectrum, we follow \cite{Respekta2025} and include in the consideration of one-photon transitions the probabilities of being empty for the initial, $1-{\cal N}_{\rm X} p_h(n_h\tilde Q_{j_h}))$, and final, $1-{\cal N}_{\rm X}p_e(n_eQ_e)$, states. ${\cal N_{\rm X}}=0$ when there is no pregenerated electron-hole pair in a NC, and ${\cal N_{\rm X}}=1$ for a NC having one pregenerated electron-hole pair. The populations $p_e(n_eQ_e)$ and $p_h(n_h \tilde Q_{j_h})$ are given by the Boltzmann distribution :
\begin{eqnarray}\label{eqpeph}
    p_{e,h}(i)=\frac{1}{Z_{e,h}}\exp\left[{-\frac{\Delta E_{e,h}(i)}{
    k_BT}}\right],   
    Z_{e,h}=\sum_{i}g_{e,h}(i)\exp\left[{-\frac{\Delta E_{e,h}(i)}{
    k_BT}}\right], \, 
\end{eqnarray}
where $k_B$ is the Boltzmann constant, $i=n_eQ_e$ and $i=n_h\tilde Q_{j_h}$ refer to the electron and hole, respectively. Energy gaps to the lowest electron and hole states are $\Delta E_e(n_eQ_e)=E_e(n_eQ_e)-E_e(1S_{e})$, $\Delta E_h(n_h\tilde Q_{j_h})=E_h(n_h\tilde Q_{j_h})-E_h(1S_{3/2})$. $Z_{e,h}$ is the canonical partition function for the electron and hole, respectively. $g_{e,h}$ is the degeneracy with respect to the total angular momentum projection of the electron and hole states, respectively. 

We consider the one-photon absorption signal  ${\cal I}_{\text{1PA}}(\hbar\omega,{\cal N}_{\rm X}) \propto  W_{\text{1PA}}(\hbar\omega,{\cal N}_{\rm X})$ with the probabilty $W_{\text{1PA}}(\hbar\omega,{\cal N}_{\rm X})$ given by the Fermi golden rule:
\begin{eqnarray}\label{eq1pa}
W_{1\text{PA}}(\hbar\omega,{\cal N_{\rm X}})=\frac{2\pi}{\hbar}\sum_{c,v} \delta[E_{cv}-\hbar\omega]\times \nonumber\\(1-{\cal N_{\rm X}}p_e(c))(1-{\cal N_{\rm X}}p_h(v))  \left|\langle c|\hat{\mathcal{V}}^{cv}_{\alpha}|v\rangle\right|^2\, .
\end{eqnarray}
Here $\hbar\omega$ is the photon energy. The summation over $c,v$ runs over all states  of the electron,  $\langle c|$, and hole, $|v\rangle$, with energies $E_e(n_eQ_e)$ and $E_h(n_h\tilde Q_{j_h})$, respectively, satisfying the energy conservation law:  $E_{cv}=E_{X}=\hbar\omega$. For  $|v\rangle$ states, the wavefunction is taken in the electron representation with $m_v=-m_h$. 
To account for the inhomogeneous broadening of the absorption spectra in the modeling of the experimental data, we replace the delta function with the Gaussian distribution function $\exp[-(E_{cv}-\hbar\omega)^2/2\Gamma^2]$ with the standard deviation $\Gamma$.   In the case ${\cal N_{\rm X}}=0$, we obtain the one-photon absorption spectrum for unexcited NCs ${\cal I}_{1\text{PA}}^0(\hbar\omega)={\cal I}_{1\text{PA}}(\hbar\omega,0)$. The differential absorption spectrum is defined as $\Delta {\cal I}_{1\text{PA}}(\hbar\omega)={\cal I}_{1\text{PA}}(\hbar\omega,0) -{\cal I}_{1\text{PA}}(\hbar\omega,1)$.

We consider two-photon absorption signal ${\cal I}_{\text{2PA}}(\hbar\omega,{\cal N}_{\rm X}) \propto W_{2\text{PA}}(2\hbar\omega)$, with $W_{2\text{PA}}(2\hbar\omega)$ calculated within the framework of the second-order perturbation theory\cite{Hutchings1992}: 
\begin{eqnarray}\label{eq2pa}
W_{2\text{PA}}(2\hbar\omega)=\frac{2\pi}{\hbar}\sum_{c,v} \delta[E_{cv}-2\hbar\omega]\times\nonumber\\\left| \sum_{c',v'}\left[\frac{\langle c| \hat{\mathcal{V}}^{cc'}_{\alpha\text{2}}|c'\rangle\langle c'|\hat{\mathcal{V}}^{c'v}_{\alpha\text{1}}|v\rangle}{E_{c'v}-\hbar\omega}+\frac{\langle c| \hat{\mathcal{V}}^{cc'}_{\alpha\text{1}}|c'\rangle\langle c'|\hat{\mathcal{V}}^{c'v}_{\alpha\text{2}}|v\rangle}{E_{c'v}-\hbar\omega}\right. \right. \nonumber\\ \left. \left.+\frac{\langle c| \hat{\mathcal{V}}^{cv'}_{\alpha\text{2}}|v'\rangle\langle v'|\hat{\mathcal{V}}^{v'v}_{\alpha\text{1}}|v\rangle}{E_{v'v}-\hbar\omega}+\frac{\langle c| \hat{\mathcal{V}}^{cv'}_{\alpha\text{1}}|v'\rangle\langle v'|\hat{\mathcal{V}}^{v'v}_{\alpha\text{2}}|v\rangle}{E_{v'v}-\hbar\omega}\right]\right|^2 \, .
\end{eqnarray}
Here, both photons have the same energy $\hbar\omega$, and the outer summation is carried out over all electron and hole states satisfying the energy conservation law $E_{cv}=E_X = 2\hbar\omega$.
In Eq.~(\ref{eq2pa}), we introduce two types of intermediate states: conduction band states, $|c'\rangle$, and valence band states, $|v'\rangle$. The energy difference between $|c'\rangle$ and $|v\rangle$ states, or between $|v'\rangle$ and $|v\rangle$ is denoted as $E_{c'v}=E_e(n'_eQ'_e)+E_h(n_h \tilde Q_{j_h})+E_g^{\text{InP}}$ or $E_{v'v}=E_h(n'_h\tilde Q'_{j'_h})-E_h(n_h \tilde Q_{j_h})$, respectively. The interband optical transition operator between the states $|c'\rangle$ and $|v\rangle$ or $|c\rangle$ and $|v'\rangle$ denoted as $\hat{\mathcal{V}}_{\alpha}^{c'v}$ or $\hat{\mathcal{V}}_{\alpha}^{cv'}$, respectively, is the same as in Eq.~(\ref{eq1pa}). For intraband transitions between the states $|c\rangle$ and $|c'\rangle$ or $|v\rangle$ and $|v'\rangle$, the optical transition operators are $\hat{\mathcal{V}}^{cc'}_{\alpha}=-e\hbar\bm{A}_{\alpha}\cdot\hat{\bm{V}}_e/c$  or $\hat{\mathcal{V}}^{v'v}_{\alpha}=-e\hbar\bm{A}_{\alpha}\cdot\hat{\bm{V}}_h/c$ , respectively, with $\hat{\bm{V}}_e$ and $\hat{\bm{V}}_h$ being the envelope velocities from Eqs.~\eqref{Eq2},\eqref{Eq3} (see ESI section S8 for details). The internal summation indices $v'$ and $c'$ pass through all intermediate states. However, in the modeling of the experimental data, we use a finite number of intermediate states listed in ESI S5. In fact, fewer intermediate states can be used since the intensity of two-photon transitions quickly reaches saturation.  As in the case of one-photon transitions, we replace the delta function in Equation~\ref{eq2pa} with a Gaussian distribution function to account for the broadening of each two-photon transition.

\section{Results and discussion}\label{results}
\begin{figure*}[h!]
 \centering \includegraphics[width=0.95\linewidth]{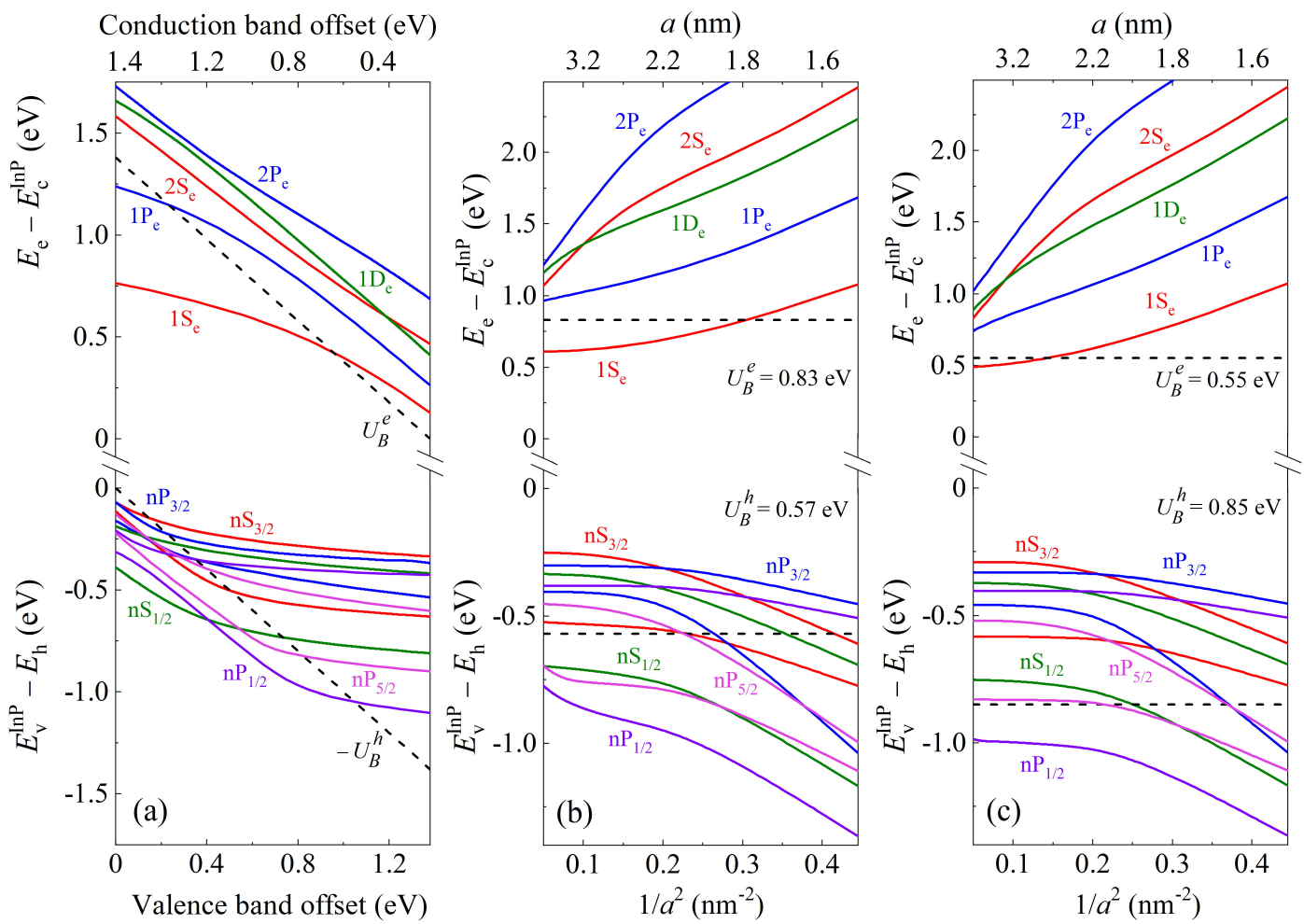}
 \caption{(a) Electron and hole energy levels given in the electron representation for a InP/ZnSe NC with $a_c=1.45$~nm and $a=4.2$~nm as functions of the band offsets $U_B^e$ and $U_B^h$ with fixed $U_B^e+U_B^h=1.4$ eV. Electron and hole energies are counted from $E_{\rm c}^{\rm InP}$ and $E_{\rm v}^{\rm InP}$, respectively. Hole states with energies above the dashed line $-U_B^h$ and electron states with energies below the dashed line $U_B^e$  are classified as localized in the InP core. Electron and hole energy levels dependencies on the total NC radius $a$ with fixed core radius $a_c=1.45$ nm for (b) the natural valence band offset $U_B^h=0.57$~eV \cite{Wei1998} and (c) valence band offset $U_B^h=0.85$ eV determined from modeling of the one- and two-photon absorption data. }
 \label{fig:Fig2}
\end{figure*}

The calculated energy spectra of non-interacting electron and hole are shown in Fig.~\ref{fig:Fig2}. The electron and hole energies are given in the electron representation and counted from the bottom of the conduction band  $E^{\text{InP}}_c$ and the top of the valence band $E^{\text{InP}}_v$ of InP, respectively. 
To calculate the hole and electron energy spectrum, we used the same set of material parameters as in Ref.~\cite{Respekta2025} (see Table S1 in ESI section S7). In the case of holes, there are many different sets of Luttinger parameters for ZnSe. We checked that other sets of Luttinger parameters \cite{Adachi2004ZnSe} result in similar hole energy spectra (see Fig. S1 in ESI section S7). We note that the energy levels and wavefunctions of the electron and hole used in Equations (\ref{eqEx}-\ref{eq2pa}) are calculated using the low-temperature parameters of InP and ZnSe (effective masses, band gaps, Kane energies), since the full set of parameters for room temperature is unknown. For the modeling of the one- and two-photon absorption spectra at room temperature, we use the room-temperature band gap of InP $E_g^{\rm InP}(T=300K)=1.34$~eV  in Eq.~\ref{eqEx} for the exciton transition energy $E_X$ and the temperature $T=300$~K in Eq.\ref{eqpeph} for populations $p_e(c)$ and $p_h(v)$.  

In Fig.~\ref{fig:Fig2}(a) we show the calculated dependencies of the electron and hole energies on the valence band and conduction band offsets in an InP/ZnSe NC with $a_c=1.45$ nm and $a=4.2$ nm. These radii correspond to InP/ZnSe NCs studied experimentally and theoretically in Ref.~\cite{Respekta2025}. For these NCs, the ground hole state is $1S_{3/2}$ over almost the entire range of valence band offsets $U_B^h$. Only at very small $U_B^h$ ($\leq 10$~meV) does the ground hole state become the $1P_{3/2}$ state, for which optical transitions to the ground electron state $1S_e$ are forbidden in the dipole approximation. However, as we show below, such small valence band offsets are not realized in the studied InP/ZnSe NCs. Recall that in the case of hole states in bare-core InP NCs, considered within the Luttinger Hamiltonian \cite{Richard1996,Efros1998}, the lowest hole state is always the $1P_{3/2}$ state, while atomistic calculations, as well as experimental studies, predict it to be  $1S_{3/2}$ \cite{Fu1998, Franceschetti1999}. Within the Luttinger model, this problem can be addressed using a general boundary conditions approach. Details of this approach, for both bare-core InP and core-shell InP/ZnSe NCs, will be published elsewhere.

\begin{figure*}[h!]
 \centering
\includegraphics[width=0.95\linewidth]{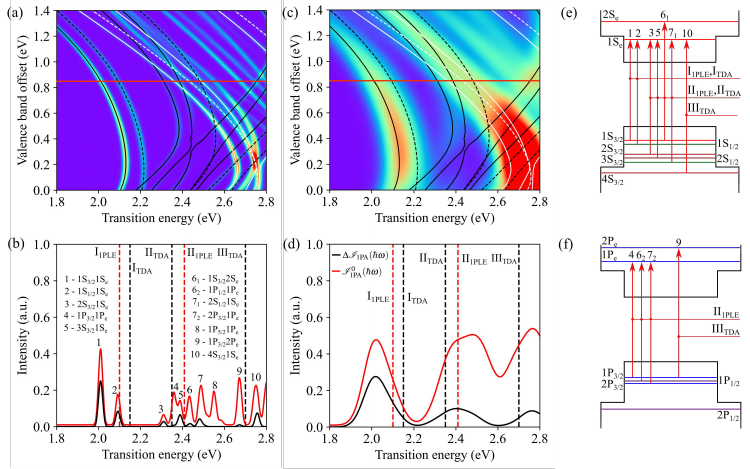}
 \caption{ The heat maps of the one-photon absorption intensity ${\cal I}^0_{\rm 1PA}(\hbar\omega)$  for different valence band offset values calculated for the inhomogeneous broadening of each exciton transition (a) $\Gamma=10$~meV  and (c) $\Gamma=50$~meV, respectively. Red lines in panels (b,d) show slices of the spectra ${\cal I}^0_{\rm 1PA}(\hbar\omega)$ from panels (a,c) for $U_B^h=0.85$~eV (horizontal red line ). Vertical red  ($\rm I_{\rm 1PLE},\rm II_{\rm 1PLE}$) and black lines ($\rm I_{\rm TDA},\rm II_{\rm TDA}, \rm III_{\rm TDA}$)  correspond to the one-photon transition energies determined in Ref.~\cite{Respekta2025} from 1PLE and differential absorption spectra, respectively. Black lines in panels (b,d) show the calculated differential absorption spectra $\Delta{\cal I}_{\rm 1PA}(\hbar\omega)$. (e,f) Schematic representation of one-photon transitions to electron states of the $S$- and $P$-symmetry, respectively. }
 \label{fig:Fig1PAnar}
\end{figure*}

Figs. \ref{fig:Fig2} (b,c) show the electron and hole energy dependencies on the total radius $a$ of the InP/ZnSe NC with fixed $a_c=1.45$~nm and fixed band offsets $U_B^{e,h}$. Fig. \ref{fig:Fig2} (b) shows the calculation results for the natural band offset $U_B^h=0.57$~eV \cite{Wei1998}. Fig. \ref{fig:Fig2}(c) shows the calculation results for the valence band offset $U_b^h=0.85$~eV, which allowed us to fit the experimental data on one- and two-photon absorption. One can see that the thickness of the ZnSe shell affects the ordering of the $1S_{3/2}$ and $1P_{3/2}$ states.  For both values of $U_B^h$, the  intersection point of $1S_{3/2}$ and $1P_{3/2}$ states in Figs.~\ref{fig:Fig2}(b,c) is close to $1/a^2\approx 0.2\ \text{nm}^{-2}$, which corresponds to the total NC radius $a=2.2$~nm or to the ZnSe shell thickness $d=0.75$ nm ($\approx 3$ monolayers). A similar intersection of the $1S_{3/2}$ and $1P_{3/2}$ states  was also observed in the numerical $k\cdot p$ calculations in Ref.\cite{Planelles2026}, where the core radius $a_c$ and the shell thickness $d$ of a InP/ZnSe NC were both varied with fixed $a=a_c+d$ and the valence band offset of $U_B^h = 0.9$ eV was used.

It is reasonable to assume that the dipole configuration at the heterointerface should be the same in NCs with a fixed core radius $a_c$, but different radii $a$. Unless the energy band offsets are additionally affected by the strain depending on the shell thickness, the calculated dependencies in Figs.~\ref{fig:Fig2}(b,c) can predict the  energy spectra of InP/ZnSe NCs with different $a$.

In the case of the electron energy spectra, shown in Figs.~\ref{fig:Fig2}(b,c), it is evident that in the entire range of the total NC radius $a$, used in calculations, only the $1S_e$ state is localized in the InP core with increasing thickness of the ZnSe shell. However, in the case of the valence band offset $U_B^h=0.85$~eV (Fig.~\ref{fig:Fig2}(c)) the electron localization energy $E_c^{\rm ZnSe}-E_e(1S_e)\approx50$~meV in the largest NCs with $a=7$~nm (ZnSe shell thickness $5.55$~nm). In this case, the energy structure of the InP/ZnSe NC is close to quasi-type II, as mentioned in Ref.~\cite{Lange2020}.  We note that using the Kane model in calculations allows us to obtain lower electron quantum size energies $E_e$, especially for excited states and for small shell thicknesses, compared to the parabolic effective mass model  (see Fig. S2 in ESI section S8).

We calculated the one-photon absorption spectra ${\cal I}_{1\text{PA}}^0(\hbar\omega)$ and ${\Delta \cal I}_{1\text{PA}}(\hbar\omega)$ as functions of the valence band offset $U_B^{h}$ and the exciton transition energy $E_X$.  Figs. \ref{fig:Fig1PAnar} (a,c) show two-dimensional heat maps of the one-photon absorption spectra ${\cal I}_{1\text{PA}}^0(\hbar\omega)$ at $T=300$ K for the inhomogeneous broadening of each exciton transition $\Gamma=10$~meV and $\Gamma=50$~meV, respectively. Figs.~\ref{fig:Fig1PAnar}(b,d) show slices of heat maps from Figs.~\ref{fig:Fig1PAnar}(a,b) for the valence band offset value $U_B^h=0.85$~eV.  
 One can see that the calculated absorption ${\cal I}_{1\text{PA}}^0(\hbar\omega)$ (red lines) and differential absorption $\Delta {\cal I}_{1\text{PA}}(\hbar\omega)$ (black lines) spectra in Fig. \ref{fig:Fig1PAnar} (b,d) are consistent with the experimental results of the one-photon photoluminescence excitation (vertical red dashed lines) and differential absorption measurements (vertical black dashed lines) from Ref.~\cite{Respekta2025}. In agreement with Ref.~\cite{Respekta2025}, we find that the lowest absorption peak of the inhomogeneously broadened ensemble is formed by the overlapping $1S_{3/2}1S_e$ and $1S_{1/2}1S_e$ transitions (Fig.~\ref{fig:Fig1PAnar}(d)).

\begin{figure*}[h!]
 \centering
 \includegraphics[width=0.95\linewidth]{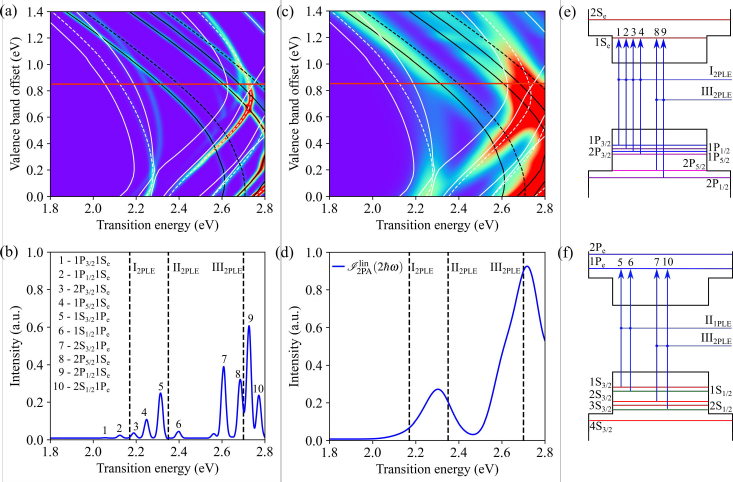}
 \caption{ The heat map of the linearly-polarized two-photon absorption intensity ${\cal I}_{\rm 2PA}^{\rm lin}(2\hbar\omega)$ for different valence band offset values, calculated for the inhomogeneous broadening of each exciton transition  (a) $\Gamma=10$~meV and (c) $\Gamma=50$~meV, respectively. Solid and dashed white lines correspond to $nP_{3/2}1S_e$ and $nP_{1/2}1S_e$ exciton energies, respectively. Solid and dashed black lines correspond to $nS_{3/2}1P_e$ and $nS_{1/2}1P_e$ exciton energies, respectively. Panels (b,d) show slices of the heat maps from panels (a,c) at the level $U_B^h=0.85$~eV (horizontal red line). Vertical dashed lines correspond to the two-photon transition energies ($\rm I_{\rm 2PLE},\rm II_{\rm 2PLE},\rm III_{\rm 2PLE}$) observed in Ref.~\cite{Respekta2025}.  (e,f) Schematic representation of two-photon transitions to electron states of the $S$- and $P$-symmetry, respectively.}
 \label{fig:Fig2PAnar}
\end{figure*}

 According to our calculations, the experimental 1PLE feature $\rm II_{1PLE}$  at the energy $\approx2.4$~eV is formed by the overlap of several exciton transitions shown in  Fig.~\ref{fig:Fig1PAnar}(b). The individual transitions   become indistinguishable when the inhomogeneous broadening is large (see Fig.~\ref{fig:Fig1PAnar}(d)).  The calculated inhomogeneously broadened one-photon absorption spectrum ${\cal I}_{1\text{PA}}^0(\hbar\omega)$ in a vicinity of $2.4$~eV has equal contributions from exciton states with symmetric  (peaks $3,5,6_1,7_1$) and anti-symmetric (peaks $4,6_2,7_2$) radial wavefunctions of the electron and hole. All of these states can contribute to the 1PLE signal. However, due to the factors $p_e(c),p_h(v)$ the differential absorption spectrum $\Delta {\cal I}_{1\text{PA}}(\hbar\omega)$ is formed from transitions involving the lowest electron state $1S_e$ or the lowest hole state $1S_{3/2}$. The third peak observed in the differential absorption measurements at $2.7$~eV \cite{Respekta2025} is formed by the $4S_{3/2}1S_e$ transition. Again, due to the factors $p_e(c),p_h(v)$, the strong $1P_{3/2}1P_e$ transition (peak 9 in Fig.~\ref{fig:Fig1PAnar}(b)) in the ${\cal I}_{1\text{PA}}^0(\hbar\omega)$ signal gives a negligible contribution to the $\Delta {\cal I}_{1\text{PA}}(\hbar\omega)$ signal. A schematic representation of one-photon transitions, responsible for the observed experimental features in Ref.~\cite{Respekta2025}, is presented in Figs.~\ref{fig:Fig1PAnar}(e,f)) for electron states of the $S$- and $P$-symmetry, respectively.
 
The calculated energy of the $1S_{3/2}1S_e$ exciton transition in Figs.~\ref{fig:Fig1PAnar}(b,d)  is about $0.1$~eV smaller compared to the energies of the lowest 1PLE and the differential absorption peaks \cite{Respekta2025}, while the higher-lying transitions fit the experimental data well. 
 If we don't include the Coulomb corrections $E_{\rm Coul}$, which are about $0.1$ eV, into the exciton transition energies, the energies of the $1S_{3/2}1S_e$ and $1S_{1/2}1S_e$ states perfectly match the positions of the lowest absorption features from Ref.~\cite{Respekta2025}  (see Fig. S3 in ESI section S9).  However, in this case, the group of transitions 3-8 from Fig.~\ref{fig:Fig1PAnar}(b) does not match the experimental feature around $\approx 2.4$~eV. 
 
 Apparently, the discrepancy in the ground exciton state energy may be a consequence of the simple model that we use, since it does not account for the finite potential barriers between the ZnSe shell and the outer matrix and considers the Coulomb interaction perturbatively. We note that in Ref.~\cite{Respekta2025} the calculated energy of the $1S_{3/2}1S_e$ transition without the Coulomb correction also matches the  lowest 1PLE peak energy, while it shifts below $2$~eV when the Coulomb correction is taken into account \cite{Planelles2026}.

\begin{figure*}[h!]
 \centering
 \includegraphics[width=1\linewidth]{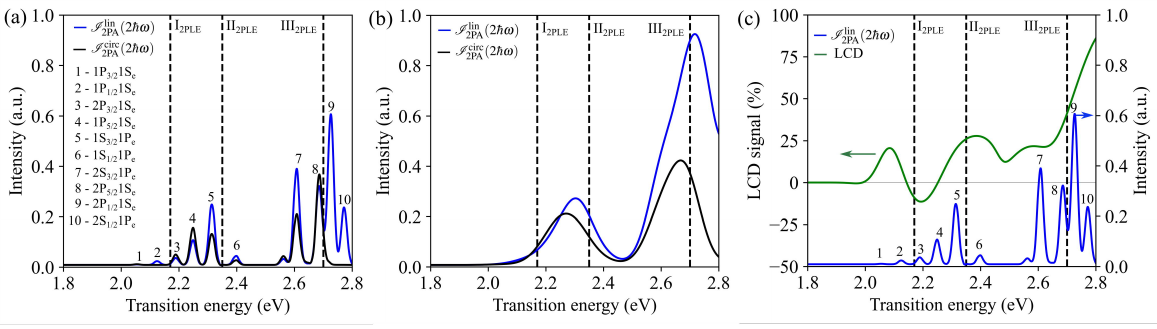}
 \caption{ The two-photon absorption spectrum in the case of linear ${\cal I}_{\rm 2PA}^{\rm lin}(2\hbar\omega)$ (blue) and circular  ${\cal I}_{\rm 2PA}^{\rm circ} (2\hbar\omega)$ (black)  ${\cal I}_{\rm 2PA}^{\rm non} (2\hbar\omega)$ polarization of photons, calculated for the inhomogeneous broadening of each exciton transition (a) $\Gamma=10$~meV and (b) $\Gamma=50$~meV, respectively. (c) Spectral dependence of the linear-circular dichroism signal for absorption spectra from panel (b) and the ${\cal I}_{\rm 2PA}^{\rm lin}(2\hbar\omega)$ spectrum calculated for the inhomogeneous broadening of each exciton transition $\Gamma=10$~meV. }
 \label{fig:Fig5}
\end{figure*}

The $U_B^h$ value determined from the modeling of the one-photon absorption spectra provides good agreement between the experimental 2PLE spectrum \cite{Respekta2025} and the two-photon absorption spectrum calculated for linear polarization of both photons $\bm e_{\alpha1}=\bm e_{\alpha2}=\bm e_x$  (see Fig.~\ref{fig:Fig2PAnar}). According to our calculations, the lowest two-photon-active exciton state $1P_{3/2}1S_e$ has a transition energy close to the lowest one-photon-active state $1S_{3/2}1S_e$. However, the $1P_{3/2}1S_e$ transition intensity is vanishing. A similar result was also obtained for CdSe NCs, where the calculated and experimentally measured intensity for $1P_{3/2}1S_e$ transition was negligible. \cite{Blanton1996}. Thus, the two-photon absorption threshold should be shifted to $2.2$~eV, which corresponds to  the $1P_{1/2}1S_e$ and $2P_{3/2}1S_e$ transitions. 

The 2PLE peak at $2.36$~eV ($\rm II_{\rm 2PLE}$ in Fig.~\ref{fig:Fig2PAnar}(b,d)), which was observed but not explained in Refs. \cite{Respekta2025}, is formed from transitions $1S_{3/2}1P_e$ and $1P_{5/2}1S_e$. We suppose that a similar 2PLE feature, observed for InP/ZnSe/ZnS NCs \cite{Tolmachev2024}, also corresponds to these transitions. A strong contribution from the $nP_{5/2}1S_e$ exciton states to the 2PLE signal was previously reported for the bare-core CdSe NCs \cite{Blanton1996}.

The high-energy peak in the 2PLE spectrum at $2.7$ eV in Ref. \cite{Respekta2025}, which was also not explained, is formed by $2S_{3/2}1P_e$, $2P_{5/2}1S_e$, $2P_{1/2}1S_e$ and $1S_{3/2}2P_e$  transitions. The dip in the calculated two-photon absorption spectrum in Fig.~\ref{fig:Fig2PAnar}(d) at $2.5$~eV is due to absence of $S$-type hole states in the energy gap between $2S_{3/2}$ and $1S_{1/2}$ states. Note, that a less pronounced dip was observed experimentally in Ref.~\cite{Respekta2025} exactly at $2.5$~eV. 

If we omit the Coulomb correction for two-photon transitions, good agreement with the experimental data from  Ref.~\cite{Respekta2025} is still observed (see Fig. S4 in ESI section S9). The $1P_{3/2}1S_e$ transition in this case is located exactly at the $2$PLE signal threshold determined in Ref.~\cite{Respekta2025}.

In Fig.~\ref{fig:Fig5} we compare the two-photon absorption intensity calculated for linearly ${\cal I}_{\rm 2PA}^{\rm lin}(2\hbar\omega)$ and circularly ${\cal I}_{\rm 2PA}^{\rm circ}(2\hbar\omega)$ polarized photons. It can be seen in Fig.~~\ref{fig:Fig5}(a) that under circularly polarized excitation, transitions $1P_{1/2}1S_e$ and $2P_{1/2}S_e$ disappear because they don't satisfy  the selection rules with angular momentum projection  $\Delta j_z=\pm2$. The linear-circular dichroism signal ${\rm LCD}=({\cal I}_{\rm 2PA}^{\rm lin}(2\hbar\omega)-{\cal I}_{\rm 2PA}^{\rm circ}(2\hbar\omega))/({\cal I}_{\rm 2PA}^{\rm lin}(2\hbar\omega)+{\cal I}_{\rm 2PA}^{\rm circ}(2\hbar\omega))$ for two-photon absorption spectra from Fig.~\ref{fig:Fig5}(b) is shown in Fig.~\ref{fig:Fig5}(c). According to our calculations, the sign reversal of the LCD signal should occur in the vicinity of the $\rm I_{\rm2PLE}$ and $\rm II_{2\rm PLE}$ features. The strong exciton transition $2P_{1/2}S_e$ for linearly polarized excitation, but forbidden for circularly polarized excitation, should lead to an increase in the LCD signal up to $75\%$ near the $\rm III_{2\rm PLE}$ experimental feature.  In Fig.S5 in ESI we compare two-photon absorption spectra for linearly polarized, circularly polarized and nonpolarized excitation, and show the linear-nonpolarized dichroism spectrum. We find that the difference between linearly polarized and nonpolarized excitation is less pronounced as compared to the difference between linearly and circularly polarized excitation.

The calculated LCD signal is determined by the fact, that the angular momentum and its projection are good quantum numbers for spherically symmetric systems. Uniaxial or tetrahedral NC shape distortion should brake spherical symmetry and decrease the LCD signal. Thus, we believe that the LCD signal may be a more reliable marker of the electron and hole Hamiltonians symmetry in InP/ZnSe NCs than the energy structure, which, as shown in Refs. \cite{Respekta2025, Planelles2026}, is almost the same for spherical and tetrahedral nanocrystals.

As was mentioned above, the our consideration does not take into account the effect of lattice strain, which requires additional complicated calculations beyond the analytical $kp$-approach. However, qualitative conclusions can be drawn about the influence of lattice strain.  According to \cite{Lange2020}, for the InP/ZnSe heterostructure, a compressive strain is present in the InP core. 
This will lead to an increase in the InP band-gap \cite{Diaz2007} by $\Delta U_{\rm str}$, but to a decrease in the total band offset $U_B^e+U_B^h$ by the same amount. In Fig.~S6 from ESI we show that to maintain a similar one-photon absorption spectrum with $\Delta U_{\rm str}=0.1$~eV, band offsets $U_B^h=0.95$~eV and $U_B^e=0.35$~eV are required.

Note that when modeling the optical spectra, we use the relationship $U_B^h+U_B^e = E_{g,\rm bulk}^{\rm ZnSe} -E_{g,\rm bulk}^{\rm InP}$. Thus, an increase in $U_B^h$ leads to the same decrease in $U_B^e$. Independently changing both band offset values would lead to a large uncertainty in $U_B^h$, since the hole energy spectrum remains virtually unchanged for $U_B^h>0.5$~eV (see Fig.~\ref{fig:Fig2}(a)). In turn, the pronounced dependence of the electron energy spectrum on the conduction band offset imposes strict restrictions on the $U_B^e$, so the key parameter that allows us to describe the experimental data is, in fact, $U_B^e$. The estimated value of $U_B^e=0.55$~eV in the absence of lattice strain is higher compared to $0.41$~eV determined in Ref.~\cite{Lange2020}. If we take into account the increase in the InP band gap by $0.1$~eV due to lattice strain, then the determined value of $U_B^e=0.35$~eV will be close to the value from Ref.~\cite{Lange2020}. 
  
A comparison of the calculated one- and two-photon absorption spectra with the experimental data allows us to conclude that the valence band offset between the InP core and the ZnSe shell is about $0.85-1$~eV, depending on the accounting for the lattice strain contribution in the calculations. This range of the valence band offsets agrees with the estimation of the lower limit for the valence band offset $0.693$~eV from Ref.~\cite{Respekta2025}. The determined band offsets in InP/ZnSe NCs indicate the predominant formation of the Zn-P bonds at the heterointerface, which should increase the natural valence band offset \cite{Jeong2022}. 
 
\section{Conclusions}

The two-photon absorption spectrum of the InP/ZnSe NCs is calculated for the first time. It explains all the experimental 2PLE features from Refs. \cite{Tolmachev2024,Respekta2025} and serves as a key tool for the estimation of the band offsets in these nanocrystals. From the comparison of the calculated absorption spectra and available experimental data, the valence band offset between InP and ZnSe of $0.85-1$~eV is estimated, in agreement with the results of Ref. \cite{Respekta2025}. This band offset range exceeds the natural valence band offset and indicates the predominant formation of the Zn-P bonds at the heterointerface. For spherical InP/ZnSe NCs we predict pronounced linear-circular dichroism, which changes sign near the two-photon absorption threshold and increases up to 90\% with increasing photon energy. The suggested experimental studies of the linear-circular dichroism signal can serve as an indicator of the symmetry of electron and hole states in InP/ZnSe nanocrystals.

\section*{Acknowledgements}
The authors thank E.L. Ivchenko, Y.M. Beltukov and D.R. Yakovlev for valuable discussions.

\section*{Supplementary Information}

\renewcommand{\thesubsection}{{S}\thesection\arabic{subsection}}

\setcounter{section}
{0}
\setcounter{equation}{0} 
\setcounter{figure}{0} 

\renewcommand{\theequation}{{S}\arabic{equation}}
\renewcommand{\thefigure}{{S}\arabic{figure}}
\renewcommand{\thetable}{{S}\arabic{table}}

\subsection{ Eight-band model for electron}\label{S1} 

The energy-band structure of the cubic semiconductors near the center of the first Brillouin zone can be well described within eight-band $\bm{k}\cdot\bm{p}$-model \cite{Pidgeon1966,BirPikus1974,Ivchenko1995}. In the bulk semiconductor, the full wave function can be presented as expansion \cite{BirPikus1974}:   
\begin{eqnarray}\label{eqS1}
\bm{\Psi}_e=\sum_{\mu=\pm1/2}\Psi^c_{\mu}(\bm{r})|S\rangle u_{\mu}+\sum_{\mu=\pm1/2}\sum_{\alpha=x,y,z}\Psi^v_{\alpha,\mu}(\bm{r})|R_{\alpha}\rangle u_{\mu},
\end{eqnarray}
where $u_{\pm1/2}$ are eigenvectors of the spin operator $\hat{\bm{S}}=1/2\hat{\bm{\sigma}}$, where $\bm{\sigma}=\{\hat{\sigma}_x,\hat{\sigma}_y,\hat{\sigma}_z\}$ are Pauli matrices, $|S\rangle$ is the Bloch function of the conduction-band edge at the $\Gamma$ point of Brillouin zone representing the eigenfunction of internal momentum $I=0$, and $|\bm{R}\rangle=\{|X\rangle,\ |Y\rangle,\ |Z\rangle\}$ is vector of the Bloch functions of the valence-band edge at the $\Gamma$ point of Brillouin zone. The combination of these functions: $1/\sqrt{2}(|R_{x}\rangle\pm|R_y\rangle)$ and $|R_z\rangle$ are the eigenfunctions of the internal momentum with $I=1$ with projections $\pm1$ and $0$ on $z$ axis, respectively \cite{BirPikus1974, Luttinger1956}. The smooth functions $\Psi^c_{\pm1/2}(\bm{r})$ are the components of the conduction band spinor $\Psi^c_e=\{\Psi_{1/2}^c(\bm{r}),\Psi^c_{-1/2}(\bm{r})\}$ and $\Psi^v_{x\pm1/2}(\bm{r})$, $\Psi^v_{y\pm1/2}(\bm{r})$, $\Psi^v_{z\pm1/2}(\bm{r})$ are $x,y,z$ components of the valence band spinor envelope vector $\bm{\Psi}^v_e=\{\Psi_x^v,\Psi^v_y,\Psi^v_z\}$ \cite{Rodina2003}. In bulk semiconductors the eight-component envelope function $\bm{\Psi}_e=\{\Psi^c_e,\bm{\Psi}^v_e\}$  is the solution of the Schr$ \ddot{\text{o}}$dinger equation:

\begin{equation}\label{eqS2}
\hat{H}_{8\times8} \left(\begin{array}{c}\Psi^c_e\\ \bm{\Psi}^{v}_e\end{array}\right)=\left(\begin{matrix}\hat{H}^c\Psi^c_e&i\frac{\hbar}{m_0}P(\hat{\bm{k}}\bm{\Psi}^v_e)\\ -i\frac{\hbar}{m_0}P(\hat{\bm{k}}\Psi^c_e)&\hat{H}^{v}\bm{\Psi}^{v}_e\end{matrix}\right)=E_{\text{e}} \left(\begin{array}{c}\Psi^c_e\\ \bm{\Psi}^{v}_e\end{array}\right).
\end{equation}
Here $E_{\text{e}}$ is counted from the bottom of the InP conduction band $E_c^{\text{InP}}$, $\hat{\bm{k}}=-i\bm{\nabla}$ is the wavevector operator. $P = -i\langle S|\hat{p}_x|X\rangle$ is the Kane matrix element, which describes the interaction between conduction and valence bands. The off-diagonal matrix element in eq. \eqref{eqS2} acts on the $\bm{\Psi}^v_e$ as a scalar product $\hat{\bm{k}}\bm{\Psi}^v_e=k_x\Psi_x^v+k_y\Psi_y^v+k_z\Psi_z^v$. The Hamiltonian for the conduction band $\hat{H}^c$, which acts on the scalar spinor $\Psi^c_e$ has the form:

\begin{eqnarray}\label{eqS3}
\hat{H}^c=\gamma_c\frac{\hbar^2\hat{k}^2}{2m_0}+U_e(\bm{r}),
\end{eqnarray}
where $\gamma_c=1+2f$ includes the contribution from remote bands $2f$ to the electron effective mass $m_e$ at the bottom of the conduction band \cite{Kane1957}. This effective mass is as:
\begin{equation}\label{eqS4}
\frac{m_0}{m_e}=\left(\gamma_c+\frac{E_p}{3}\left[\frac{2}{E_g}+\frac{1}{(E_g+\Delta_{\text{SO}})}\right]\right).
\end{equation}
In equation \eqref{eqS4} \(E_p = 2P^2/m_0\) is the 
Kane energy, $E^{\text{InP}}_g=E^{\text{InP}}_c-E^{\text{InP}}_v$ is the band gap of InP ($E^{\text{InP}}_v$ — the top of the InP valence band) and $\Delta_{\text{SO}}$ is the spin-orbit splitting of the valence band. The full eight-band $\bm{k}\cdot\bm{p}$ model valence-band Hamiltonian $\hat{H}^v$ consists of six-band Luttinger Hamiltonian \cite{Luttinger1956} in the electron representation. We neglect the $\propto k^2$ terms in the valence band part of the Hamiltonian $\hat{H}^v$ and present it in the spherical approximation: 
\begin{eqnarray}\label{eqS5}
\hat{H}^{v}=-E_{g}^{\text{InP}}+\frac{1}{3}\Delta_{\text{SO}}\left[(\bm{I}\bm{\sigma})-1\right]+U_e(\bm{r})\,.
\end{eqnarray}
In this representation the Hamiltonian $\hat{H}^v$ should be considered as the $2\times2$ matrix acting on the spinor vector $\bm{\Psi}^v_e$ rather than the $6\times6$ matrix Hamiltonian acting on the six component wave function as in \cite{Luttinger1956}. Correspondingly $\hat{\bm{I}} = \{\hat{I}_x, \hat{I}_y, \hat{I}_z\}$ in eq. \eqref{S5} is the vector operator of inner orbital momentum of the valence band  $(I=1)$. Both eq. \eqref{S3} and eq. \eqref{S5} include a smooth step-like spherically symmetric potential $U_e(\bm{r})$:
\begin{eqnarray}\label{eqS6}
U_{e}(r)=\begin{cases}0 & r < a_{c}\\U_{B}^{e} & a_c\leq r\leq a\\+\infty & r > a\end{cases}.
\end{eqnarray}

The valence band contribution $\bm{\Psi}^v_e$ to the full electron wave function $\bm{\Psi}_e$ can be expressed through $\Psi^c_e$ as:
\begin{eqnarray}\label{eqS7}
\bm \Psi^v_e= -\frac{i\hbar P}{m_0 E_p}\left(\frac{m_0}{m_c(E_e,U_e)}-\gamma_c\right)(\bm k \Psi^c_e)-\frac{\hbar P}{2m_0E_p}(g_c(E_e,U_e)-g_0)[\bm \sigma \times \bm k]\Psi^c_e.
\end{eqnarray}

In eq. \eqref{eqS7} energy-dependent electron mass
$m_c(E_e)$ equals:
\begin{equation}\label{eqS8}
\frac{m_0}{m_c(E_e,U_e)}=\left(\gamma_c+\frac{E_p}{3}\left[\frac{2}{(E_e-U_e+E_g)}+\frac{1}{(E_e-U_e+E_g+\Delta_{\text{SO}})}\right]\right)\, .
\end{equation}
The expression for  $g$-factor in equation \eqref{eqS7} is:
\begin{equation}\label{eqS9}
g_c(E_e,U_e)=g_0-\frac{2E_p}{3}\frac{\Delta_{\text{SO}}}{(E_e-U_e+E_g)(E-U_e+E_g+\Delta_{\text{SO}})}\,
\end{equation}

It need to be noticed that the second term in eq. \eqref{eqS7} can be considered as perturbation. As result, Shr$\ddot{\text{o}}$odinger-like equation \eqref{eqS10} can be obtained by substituting first term of eq. \eqref{eqS7} into \eqref{eqS2}: 

\begin{equation}\label{eqS10}
\frac{\hbar^2\hat{k}^2}{2m_c(E_e,U_e)}\Psi^c_e+U_e(\bm{r})\Psi^c_e=E_e\Psi^c_e.
\end{equation}
The wave function normalization condition of the full electron wave function $\bm{\Psi}_e$ has the form:
\begin{equation}\label{eqS11}
\int\displaylimits_{\mathbb{R}^3}\left(|\Psi^{c}_e|^2+|\mathbf{\Psi}^v_e|^2 \right)\mathrm{d}^3\bm{r}=1\, .
\end{equation}

\subsection{Luttinger Hamiltonian}\label{S2}
Six-band $\bm{k}\cdot\bm{p}$ Luttinger model \cite{Luttinger1956} for electron in the valence band in the hole representation is discussed in this section. The six-band Schr$\ddot{\text{o}}$dinger equation has the form:
\begin{eqnarray}\label{eqS12}
\hat{H}_{6\times6}\bm{\Psi}_h=\frac{\hbar^2}{2m_0}\left[(\gamma_1+4\gamma_2)k^2-6\gamma_2\sum_{\alpha=x,y,z}k_{\alpha}^2I^2_{\alpha}-12\gamma_3\sum_{\alpha\neq\beta}\{k_{\alpha}k_{\beta}\}\{I_{\alpha}I_{\beta}\}\right]\bm{\Psi}_h\nonumber\\-\frac{1}{3}\Delta_{\text{SO}}[(\bm{I}\bm{\sigma})-1]\bm{\Psi}_h+U_h(r)\bm{\Psi}_h=E_h\bm{\Psi}_h.
\end{eqnarray}
Where,  $\bm{\Psi}_h=\{\Psi_x^h,\Psi^h_y,\Psi^h_z\}$  is the vector spinor. 
We neglect cubic symmetry of the hole energy dispersion, so we can apply spherically-symmetric approximation. In spherically symmetric approximation with $\gamma=(2\gamma_2+3\gamma_3)/5$ and Luttinger parameters $\gamma_1,\ \gamma_2,\ \gamma_3$ \cite{Luttinger1956}. For representation \eqref{eqS12} the relation  $(\bm{I}\bm{T})\bm{\Psi}_h=i[\bm{T}\times\bm{\Psi}_h]$ can be applied, where $\bm{T}$ is arbitrary vector. This property and knowledge of $\gamma$ allows us to re-write eq. \eqref{eqS12} in vector form:


\begin{eqnarray}\label{eqS13}
\hat{H}_{6\times6}\bm{\Psi}_h=\frac{\hbar^2k^2}{2m_0}\beta_h\bm{\Psi}_h+\frac{\hbar^2}{2m_0}[\beta_l-\beta_h]\bm{k}(\bm{k}\cdot \bm{\Psi}_h)-\nonumber\\-\frac{i\Delta_{\text{SO}}}{3}[\bm{\sigma}\times \bm{\Psi}_h]+\frac{\Delta_{\text{SO}}}{3}\bm{\Psi}_h+U_h(\bm{r})\bm{\Psi}_h=E_h\bm{\Psi}_h,
\end{eqnarray}
where  parameters $\beta_h=\gamma_1-2\gamma$, $\beta_l=\gamma_1+4\gamma$. The normalization condition for the hole wave function $\bm{\Psi}_h$ is given by $\int\displaylimits_{\mathbb{R}^3}|\mathbf{\Psi}_h|^2\mathrm{d}^3\bm{r}=1\, .$ Equations \eqref{eqS12} and \eqref{eqS14} have include a step-like spherically-symmetric potential $U_h(r)$:
\begin{eqnarray}\label{eqS14}
 \ U_h(r)=\begin{cases}0 & r \leq a_c \\U_B^h & a_c<r\leq a\\+\infty & r > a\end{cases}
.
\end{eqnarray}

\subsection{Boundary conditions at the interface}\label{S3}

We define the velocity operator in vector form as $\hat{\bm{V}}=\frac{1}{\hbar}\frac{\partial \hat{H}}{\partial \bm{k}}$. The inner product of the $n$-component ($n=8$ and $n=6$ for electron Kane and six-band Luttinger models, respectively) wave functions considered as $(\bm{\Psi}_{\alpha},\bm{\Psi}_{\beta})=\Psi_{\alpha,n}^{*}\Psi_{\beta,n}$, where $\bm{\Psi}_{\alpha}\equiv\bm{\Psi}_{\alpha}(\bm{r})$ and $\bm{\Psi}_{\beta}\equiv\bm{\Psi}_{\beta}(\bm{r})$ are two arbitrary solutions of  the Schr$\ddot{\text{o}}$dinger equation, defined in each volume layer and corresponding to  $\varepsilon_{\alpha}$ and $\varepsilon_{\beta}$ energies. This allows us to introduce the matrix of density probability flux as:

\begin{equation}\label{eqS15}
\bm{J}^{\alpha,\beta}=\frac{1}{2}\left[\left(\bm{\Psi}_{\alpha},\hat{\bm{V}}\bm{\Psi}_{\beta}\right)+\left(\hat{\bm{V}}\bm{\Psi}_{\alpha},\bm{\Psi}_{\beta}\right)\right].
\end{equation}
For the eight-band Hamiltonian \eqref{eqS2} we can write the velocity operator as: 

\begin{equation}\label{eqS16}
\hat{\bm{V}}_e=\frac{1}{\hbar}\frac{\partial \hat{H}_{8\times8}}{\partial \bm{k}}=\begin{bmatrix}\frac{\gamma_c\hbar}{m_0}\hat{\bm{k}} & i\frac{P}{m_0}\bm{e}_x& i\frac{P}{m_0}\bm{e}_y& i\frac{P}{m_0}\bm{e}_z \\-i\frac{P}{m_0}\bm{e}_x & & & \\-i\frac{P}{m_0}\bm{e}_y & & \frac{1}{\hbar}\frac{\partial \hat{H}^v}{\partial \bm{k}}& \\-i\frac{P}{m_0}\bm{e}_z & & &  \end{bmatrix},
\end{equation}
where $\bm{e}_{x,y,z}$ are Cartesian unit orts. In the case of eight-band $\bm{k}\cdot\bm{p}$ Kane model ($\gamma_c\neq0$ and $\gamma_1=\gamma=0$) as in eq. \eqref{eqS2} valence band Hamiltonian $\hat{H}^v$ \eqref{eqS5} doesn't depend on $\bm{k}$, thus $\frac{1}{\hbar}\frac{\partial \hat{H}^v}{\partial \bm{k}}=0$. Using electron velocity \eqref{eqS16} allows us to write the flux at the heterointerface for the eight-band Kane model with dispersion for electrons only:
\begin{eqnarray}\label{eqS17}
\bm{J}^{\alpha,\beta}_e=\frac{1}{2}\left[\frac{\gamma_c\hbar}{m_0}\left((\Psi^c_{\alpha})^{\dagger}\hat{\bm{k}
}\Psi^c_{\beta}+(\hat{\bm{k}}\Psi^c_{\alpha})^{\dagger}\Psi^c_{\beta}\right)+i\frac{2P}{m_0}\left((\Psi^c_{\alpha})^{\dagger}\bm{\Psi}_{\beta}^{v}-(\bm{\Psi}_{\alpha}^v)^{\dagger}\Psi^c_{\beta}\right)\right].
\end{eqnarray}
However, the dependence between $\Psi^c_e$ and $\bm{\Psi}^v_e$, which defined eq. \eqref{eqS7} allows us to rewrite \eqref{eqS17} in a form, that depends only on the spinor component:
\begin{eqnarray}\label{eqS18}
\bm{J}_e^{\alpha,\beta}=\frac{1}{2}\left[\frac{\hbar}{m_0}\left(\Psi^{c,\dagger}_{\alpha}\frac{m_0}{m_c(E_e,U_e)}(\hat{\bm{k}}\Psi_{\beta}^{c})+\frac{m_0}{m_c(E_e,U_e)}(\hat{\bm{k}}\Psi_{\alpha}^{c})^{\dagger}\Psi_{\beta}^{c}\right)\right.\nonumber\\\left.-\frac{\hbar}{2m_0}(g_c(E_e,U_e)-g_0)\left(\Psi^{c,\dagger}_{\alpha}(i[\bm{\sigma}\times\bm{\hat{k}}]\Psi^c_{\beta})+(i[\bm{\sigma}\times\bm{\hat{k}}]\Psi^c_{\alpha})^{\dagger}\Psi^c_{\beta}\right)\right].
\end{eqnarray}
Let us define the full velocity for the eight-band Kane model in vector form $\bm{V}_e$: 
\begin{eqnarray}\label{eqS19}
\bm{V}_e=\hat{\bm{V}}_e\bm{\Psi}_e=\frac{\hbar}{m_c(E_e,U_e)}\hat{\bm{k}}\Psi^c_e-\frac{i\hbar}{2m_0}\left[g_c(E_e,U_e)-g_0\right][\bm{\sigma}\times\hat{\bm{k}}]\Psi^c_e.
\end{eqnarray}
Standard boundary conditions (BCs) for electron defined as $\Psi^c=\text{const}$ and $(\bm{\tau}\cdot \bm{J}_e^{\alpha,\beta})=\text{const}$ at the boundary. For normal vector to the spherical boundary $\bm{\tau}=\bm{r}/|\bm{r}|$, the normal components of $\bm{J}^{\alpha,\beta}_e$ and $\bm{V}_e$ have form: 

\begin{eqnarray}\label{eqS20}
\bm{\tau}\cdot\bm{J}_e^{\alpha,\beta}=\frac{1}{2}\left[\frac{\hbar}{m_0}\left(\Psi^{c,\dagger}_{\alpha}\frac{m_0}{m_c(E_e,U_e)}(-i\frac{\partial }{\partial r}\Psi_{\beta}^{c})+\frac{m_0}{m_c(E_e,U_e)}(-i\frac{\partial }{\partial r}\Psi_{\alpha}^{c})^{\dagger}\Psi_{\beta}^{c}\right)\right.\nonumber\\\left.-\frac{\hbar}{2m_0r}(g_c(E_e,U_e)-g_0)\left(\Psi^{c,\dagger}_{\alpha}(i\bm{r}\cdot[\bm{\sigma}\times\bm{\hat{k}}]\Psi^c_{\beta})+(i\bm{r}\cdot[\bm{\sigma}\times\bm{\hat{k}}]\Psi^c_{\alpha})^{\dagger}\Psi^c_{\beta}\right)\right].
\end{eqnarray}

\begin{eqnarray}\label{eqS21}
\bm{\tau} \cdot\bm{V}_e=-\frac{i\hbar}{m_c(E_e,U_e)}\frac{\partial }{\partial r}\Psi^c_e-\frac{i\hbar}{2m_0r}\left[g_c(E_e,U_e)-g_0\right](\bm{r}\cdot[\bm{\sigma}\times\hat{\bm{k}}])\Psi^c_e.
\end{eqnarray}
The last term in eq. \eqref{eqS21} can be considered within the framework of the perturbation theory \cite{Rodina2003}. The full and normal velocities for electron have form:


\begin{equation}\label{eqS22}
\bm{V}_e= \hat{\bm{V}}_e\bm{\Psi}_e\approx\frac{\hbar\hat{\bm{k}}\Psi^c_e}{m_c(E_e,U_e)}, \ V^e_{\bm{\tau}}=-\frac{i\hbar}{m_c(E_e,U_e)}\frac{\partial }{\partial r}\Psi^c_e \, .
\end{equation}


The vector operator  $\hat{\bm{V}}_{h}=\frac{1}{\hbar}\frac{\partial \hat{H}_{6\times6}}{\partial \bm{k}}=\{\hat{V}_x^{h}, \hat{V}_y^{h}, \hat{V}^{h}_z\}$ is the canonical velocity of the six-band Luttinger Hamiltonian $\hat{H}_{6\times6}$. Each component, which must be applied to $\bm{\Psi}_h$, are given in eqs. \eqref{eqS23}-\eqref{eqS25}:

\begin{eqnarray}\label{eqS23}
\hat{V}_{x}^{h}=\frac{\hbar}{m_0}\beta_h\begin{bmatrix}\hat{k}_x & 0& 0 \\0 & \hat{k}_x& 0\\0 & 0& \hat{k}_x \end{bmatrix}+\frac{\hbar}{m_0}\frac{(\beta_l-\beta_h)}{2}\begin{bmatrix}2\hat{k}_x & \hat{k}_y& \hat{k}_z \\\hat{k}_y & 0& 0\\\hat{k}_z & 0& 0 \end{bmatrix}
\end{eqnarray}


\begin{eqnarray}\label{eqS24}
\hat{V}_{y}^{h}=\frac{\hbar}{m_0}\beta_h\begin{bmatrix}\hat{k}_y & 0& 0 \\0 & \hat{k}_y& 0\\0 & 0& \hat{k}_y \end{bmatrix}+\frac{\hbar}{m_0}\frac{(\beta_l-\beta_h)}{2}\begin{bmatrix}0 & \hat{k}_x& 0\\\hat{k}_x & 2\hat{k}_y& \hat{k}_z\\0 & \hat{k}_z& 0 \end{bmatrix}
\end{eqnarray}


\begin{eqnarray}\label{eqS25}
\hat{V}_{z}^{h}=\frac{\hbar}{m_0}\beta_h\begin{bmatrix}\hat{k}_z & 0& 0 \\0 & \hat{k}_z& 0\\0 & 0& \hat{k}_z \end{bmatrix}+\frac{\hbar}{m_0}\frac{(\beta_l-\beta_h)}{2}\begin{bmatrix}0 & 0& \hat{k}_x\\0 & 0& \hat{k}_y\\\hat{k}_x & \hat{k}_y& 2\hat{k}_z \end{bmatrix}
\end{eqnarray}

Projecting these components onto the normal vector $\bm{\tau}$ leads to the canonical normal velocity operator $\hat{V}^{h}_{\bm{\tau}}=(\bm{\tau}\cdot \hat{\bm{V}}_{h})$:

\begin{equation}\label{eqS27}
\hat{V}^{h}_{\bm{\tau}}=\frac{\hbar}{m_0}\frac{(\beta_l-\beta_h)}{2}\begin{bmatrix}2\tau_x\hat{k}_x &\tau_x\hat{k}_y + \tau_y\hat{k}_x & \tau_x\hat{k}_z + \tau_z \hat{k}_x \\\tau_x\hat{k}_y + \tau_y\hat{k}_x&2\tau_y\hat{k}_y\ & \tau_y\hat{k}_z + \tau_z\hat{k}_y\\\tau_x\hat{k}_z + \tau_z \hat{k}_x  & \tau_y\hat{k}_z + \tau_z\hat{k}_y & 2\tau_z\hat{k}_z \end{bmatrix}+\frac{\hbar}{m_0}\beta_h ({\bm \tau}\cdot {\hat{\bm k}}) \begin{bmatrix}1& 0& 0 \\0 & 1& 0\\0 & 0& 1 \end{bmatrix}.
\end{equation}

Thus, standard BCs for holes has form: $\bm{\Psi}_h=\text{const}\ ,\bm{V}^{h}_{\bm{\tau}}=\hat{V}_{\bm{\tau}}^{h}\bm{\Psi}_h=\text{const}.$ It necessary to  notice, that  standard BCs are canonical in case of hole wave function $\bm{\Psi}_h$.

\subsection{Analytical expressions for the eight-band wave function in spherically symmetric potential }\label{S4}
For the electron $\bm{\Psi}_e$ and hole $\bm{\Psi}_h$ wave functions in a potential with spherical symmetry, we apply a basis of Bloch wave functions $u^v_{3/2,\mu}(\mu=\pm3/2,\ \pm1/2)$ with full inner momentum $J=3/2$ in the valence band $\Gamma_8$ and $u^v_{1/2,\mu}(\mu=\pm1/2)$ with $J=1/2$ in the valence band $\Gamma_7$. Bloch functions of inner momentum can be constructed using the rule of angular momentum coupling $\bm{J}=\bm{I}+\bm{S}$. We use in this article Bloch basis functions as they are defined in \cite{Ivchenko2005}.

The electron wave function $\bm{\Psi}_e$  in the eight-band $\bm{k}\cdot\bm{p}$ Kane model, as in Ref.  \cite{Rodina2003}, can be written in the new basis of Bloch functions as:
\begin{eqnarray}\label{eqS28}
\bm{\Psi}_e=\sum_{\mu=\pm1/2}\Psi^c_{\mu}(\bm{r})u^c_{1/2,\mu}+\sum_{\mu=\pm3/2,\pm1/2}\Psi^v_{\mu}(\bm{r})u^v_{3/2,\mu}+\sum_{\mu=\pm1/2}\Psi^{\text{SO}}_{\mu}u^{\text{SO}}_{1/2,\mu}.
\end{eqnarray}

In a spherically symmetric potential, all states of electrons and holes are characterized by the total angular momentum $\bm{j}=\bm{J}+\bm{L}$  and it's projection $m=j_z$. Thus, electron wave function as was said previously in \ref{S3} characterized not only by $j_e$, but with projection $m_e=j^e_{z}$. 
Choice of sign $\pm$ in this case corresponds to ratio $j_e=l_e\pm1/2$, where "even" and "odd" values of orbital angle moment of envelope wave function in conduction band $l_e=0,1...$ relates to even $+$ and odd $-$ states of electrons, which envelope wave function in conduction band has form:
\begin{eqnarray}\label{eqS29}
\Psi^{c,\pm}_{j_e,m_e}(\bm{r})=R_{l_e}^{c,\pm}(\bm{r})\Omega^e_{j_e,l_e=j_e\mp1/2,m_e}(\theta,\varphi).
\end{eqnarray}
Spherical spinors $\Omega_{j,l,m}(\theta,\varphi)$ constructed by rule of angle coupling as:
\begin{eqnarray}\label{eqS30}
\Omega_{j,l,m}(\theta,\varphi)=\left(\begin{array}{c}C^{j,m}_{1/2,1/2,l,m-1/2}Y_{l,m-1/2}(\theta,\varphi)\\ C^{j,m}_{1/2,-1/2,l,m+1/2}Y_{l,m+1/2 }(\theta,\varphi)\end{array}\right),
\end{eqnarray}
$C^{j,m}_{j_1,m_1,j_2,m_2}$ - Clebsh-Gordan's coefficients and $Y_{l,m}$ - shperical harmonics \cite{Edmonds1957}. Hovewer, as was mentioned in \ref{S3}, we neglect non-degeneracy on $j_e$ of electron energy $E_{\text{e}}$ with the same value of $l_e$.

The six-band $\bm{k}\cdot\bm{p}$-Luttinger Hamiltonian doesn't count interaction between conduction band and valence band. In this case the total wave function of hole $\bm{\Psi}_h$, which characterized by the total angular momentum $j_h$ and it's projection $m_h=j_z^h$, expands only on valence band basis functions:
\begin{eqnarray}\label{eqS31}
\bm{\Psi}_h=\sum_{\mu=\pm3/2,\pm1/2}\Psi^h_{\mu}(\bm{r})u^v_{3/2,\mu}+\sum_{\mu=\pm1/2}\Psi^{h_\text{SO}}_{\mu}u^{\text{SO}}_{1/2,\mu},
\end{eqnarray}

which allows to rewrite spinor vector function $\bm{\Psi}_h$ in form of new envelope wave functions of valence band:
\begin{eqnarray}\label{eqS32}
\Psi_h=\left(\begin{array}{c}\Psi^h_{3/2}\\ \Psi^h_{1/2}\\ \Psi^h_{-1/2}\\ \Psi^h_{-3/2}\end{array}\right), \ \Psi^{\text{SO}}_h=\left(\begin{array}{c}\Psi^{h_\text{SO}}_{1/2}\\ \Psi^{h_\text{SO}}_{-1/2}\end{array}\right)
\end{eqnarray}

Four-component hole wave function $\Psi^{h,\pm}_{j_h,m_h}(\bm{r})$ includes contributions of two orbital moments $l_h$ and spin-orbital hole wave function $\Psi^{h_\text{SO},\pm}_{j_h,m_h}(\bm{r})$ characterized by orbital momentum $l_{\text{SO}}$:
\begin{eqnarray}
\Psi^{+}_{j_h,m_h}(\bm{r})=R^{+}_{l_{h_1}=j_h+1/2}\Omega^v_{j_h,l_{h_1}=j_h+1/2,m_h}(\theta,\varphi)+R^{+}_{l_{h_2}=j_h-3/2}\Omega^v_{j_h,l_{h_1}=j_h-3/2,m_h}(\theta,\varphi),\label{eqS33}\\
\Psi^{-}_{j_h,m_h}(\bm{r})=R^{-}_{l_{h_1}=j_{h}-1/2}\Omega^v_{j_h,l_{h_1}=j_h-1/2,m_h}(\theta,\varphi)+R^{-}_{l_{h_2}=j_h+3/2}\Omega^v_{j_h,l_h=j+3/2,m_h}(\theta,\varphi),\label{eqS34}\\
\Psi^{\text{SO},\pm}_{j_h,m_h}(\bm{r})=R^{\pm}_{l_{\text{SO}}=j_h\pm1/2}\Omega^{\text{SO}}_{j_h,l_{\text{SO}}=j_h\pm1/2,m_h}(\theta,\varphi).\label{eqS35}
\end{eqnarray}
Spherical spinors $\Omega^{\text{SO}}_{j_h,l_{\text{SO}}=j_h\pm1/2,m_h}$ defined as eq. \eqref{eqS30} and four-component spherical functions have form:
\begin{eqnarray}
\Omega^v_{j,l,m}=\left(\begin{array}{c}C^{j,m}_{3/2,3/2,l,m-3/2}Y_{l,m-3/2}\\ C^{j,m}_{3/2,1/2,l,m-1/2}Y_{l,m-1/2}\\ C^{j,m}_{3/2,-1/2,l,m+1/2}Y_{l,m+1/2}\\ C^{j,m}_{3/2,-3/2,l,m+3/2}Y_{l,m+3/2}\end{array}\right).\label{eqS36}
\end{eqnarray}
Radial wave functions of even $R^+_{l_{\alpha}}(r)$ and odd $R^-_{l_{\alpha}}(r)$ hole states ($\alpha: h_1,\ h_2, \ \text{SO}$), normalization equations for radial components of electron and hole wave functions are described in S5 in detail. We also must notice, that for calculation of all transitions, we need connection between hole and valence electron $\bm{\Psi}_v$ wave function. It  can be described by time-inversion (T-symmetry) operator : $\hat{K}=-i\hat{\sigma}_y\mathbb{C}$, where $\mathbb{C}$ is complex conjugation which leads to $\bm{\Psi}_v=\hat{K}\bm{\Psi}_h$
\subsection{Analytical radial wave-functions for charge carriers}\label{S5}
\subsubsection{Electrons}
Radial component $R_{l_e}^c(r)$ of electron conduction band envelope wave function is piecewise:


\begin{equation}\label{eqS37}
R^c_{l_e}(r)=\begin{cases}R^A_{l_e} & r \leq a_c \\R^B_{l_e} & a_c < r\leq a  \,  \end{cases},  
\end{equation}


which has direct connection with radial component of valence band contribution $\mathcal{R}_{l_e}^v(r)$ (discussed in details in \cite{Rodina2003}). Material parameters $m_e\equiv m_e(r)$, $E_p\equiv E_p(r)$ and $\Delta_{\text{SO}}\equiv \Delta_{\text{SO}}(r)$ also are piecewise functions:
\begin{eqnarray}\label{eqS38}
m_e(r)=\begin{cases}m^{\text{InP}}_e &  r \leq a_c\\m_e^{\text{ZnSe}} & a_c < r\leq a\end{cases} \ E_p(r)=\begin{cases}E_p^{\text{InP}} & r \leq a_c\\E_p^{\text{ZnSe}} & a_c < r\leq a\end{cases} \ \Delta_{\text{SO}}(r)=\begin{cases}\Delta_{\text{SO}}^{\text{InP}} & r \leq a_c\\ \Delta_{\text{SO}}^{\text{ZnSe}} & a_c < r\leq a\end{cases} 
\end{eqnarray}

Components $R_{l_e}^A$ and $R_{l_e}^B$ are described below, containing constants $A_{l_e}, \ B_{l_e}$. These constants can be  found from boundary conditions analytically.

\begin{equation}\label{eqS39}
\begin{array}{c}R_{l_e}^A=\mathcal{C}_{l_e}j_{l_e}(k_{l_e}r) \, , \\ R_{l_e}^B=\mathcal{C}_{l_e}[A_{l_e}h_{l_e}^{(1)}(i\varkappa_{l_e}r)+B_{l_e}h_{{l_e}}^{(2)}(i\varkappa_{l_e}r)],\\ \end{array}
\end{equation}
where wave numbers $k_{l_e}$ and $\varkappa_{l_e}$ related the electron level energy $E_{\text{e}}$ as:
\begin{equation}\label{eqS40}
E_{\text{e}}=\frac{\hbar^2k_{l_e}^2}{2m_c(E_e,0)}; \ U^e_B-E_{\text{e}}=\frac{\hbar^2\varkappa_{l_e}^2}{2m_c(E_e,U_B^e)},
\end{equation}
where $j_l(k_{l_e}r)$, $h^{(1)}(i\varkappa_{l_e}r)$ and $h^{(2)}(i\varkappa_{l_e}r)$ - Bessel and Hankel functions \cite{Abramowitz1972}. Constant $\mathcal{C}_{l_e}$ can be found from the radial wave function normalization condition:

\begin{align}
\int\displaylimits_{0}^{+\infty}|R^c_{l_e}|^2r^2\mathrm{d}r+\int\displaylimits_{0}^{+\infty}|\mathcal{R}^v_{l_e}|^2r^2\mathrm{d}r=1,\label{eqS41} \\ |\mathcal{R}^v_{l_e}|^2=\frac{\hbar^2}{2m_0E_p}\left(\frac{m_0}{m_c(E_{\text{e}},U_e)}-\gamma_{c}\right)^2\left(\left|\frac{\partial R^c_{l_e}}{\partial r}\right|^2+\frac{l_e(l_e+1)}{r^2}|R^c_{l_e}|^2\right)\, . \label{eqS42}
\end{align}

\subsubsection{Holes}
Radial components of hole wave functions are piecewise:

\begin{equation}\label{eqS43}
R_{l_{h_1}}^{\pm}(r),\  R_{l_{h_2}}^{\pm}(r), \  R_{l_\text{SO}}^{\pm}(r)=\begin{cases}R^{A,\pm}_{l_{h_1}}, \ R^{A,\pm}_{l_{h_2}}, \ R^{A,\pm}_{l_\text{SO}}& r \leq a_c\\ R^{B,\pm}_{l_{h_1}}, \ R^{B,\pm}_{l_{h_2}}, \ R^{B,\pm}_{l_\text{SO}} & a_c<r\leq a\end{cases} \ .
\end{equation}
Luttinger parameters $\gamma_1\equiv\gamma_1(r)$ and $\gamma\equiv\gamma(r)$ are also piecewise functions: 


\begin{eqnarray}\label{eqS44}
\gamma(r)=\begin{cases}\gamma^{\text{InP}} & r \leq a_c\\\gamma^{\text{ZnSe}} & r>a_c\end{cases}, \ \gamma_1(r)=\begin{cases}\gamma_1^{\text{InP}} & r \leq a_c\\\gamma_1^{\text{ZnSe}} & r>a_c\end{cases}
\end{eqnarray}
Radial components of even ("+") functions in InP core have form: 

\begin{align}
R^{A,+}_{l_{h_1}}=\mathcal{C}_{j_h}^+\{-c_1j_{j_h+1/2}(k_{lh}r)-c_2j_{j_h+1/2}(k_{hh}r)-c_3j_{j_h+1/2}(k_{sh}r)\}\label{eqS45}\\
R^{A,+}_{l_{h_2}}=\mathcal{C}_{j_h}^+\{-c_1\sqrt{\frac{3(2j_h-1)}{(2j_h+3)}}j_{j_h-3/2}(k_{lh}r)+c_2\sqrt{\frac{(2j_h+3)}{3(2j_h-1)}}j_{j_h-3/2}(k_{hh}r)-c_3\sqrt{\frac{3(2j_h-1)}{(2j_h+3)}}j_{j_h-3/2}(k_{sh}r)\}\label{eqS46}\\
R^{A,+}_{l_\text{SO}}=\mathcal{C}^+_{j_h}\{c_1\sqrt{\frac{j_h}{(2j_h+3)}}\frac{2\varepsilon-(\gamma_1+2\gamma)k_{lh}^2}{\gamma k_{lh}^2}j_{j_h+1/2}(k_{lh}r)+c_3\sqrt{\frac{j_h}{(2j_h+3)}}\frac{2\varepsilon-(\gamma_1+2\gamma)k_{sh}^2}{\gamma k_{sh}^2}j_{j_h+1/2}(k_{sh}r)\}\label{eqS47}
\end{align}

Radial components of odd ("-") functions in InP core have form:

\begin{align}
R_{l_{h_1}}^{A,-}=\mathcal{C}^-_{j_h}\{c_4\sqrt{2j_h-1}j_{j_h-1/2}(k_{lh}r)+c_5\sqrt{2j_h-1}j_{j_h-1/2}(k_{hh}r)+c_6\sqrt{2j_h-1}j_{j_h-1/2}(k_{sh}r)\}\label{eqS48}\\
R_{l_{h_2}}^{A,-}=\mathcal{C}^-_{j_h}\{-c_4\sqrt{3(2j_h+3)}j_{j_h+3/2}(k_{lh}r)+c_5\frac{(2j_h-1)}{\sqrt{3(2j_h+3)}}j_{j_h+3/2}(k_{hh}r)-c_6\sqrt{3(2j_h+3)}j_{j_h+3/2}(k_{sh}r)\}\label{eqS49}\\
R_{l_\text{SO}}^{A,-}=\mathcal{C}^-_{j_h}\{-c_4\sqrt{j_h+1}\frac{(\gamma_1+2\gamma)k_{lh}^2-2\varepsilon}{\gamma k_{lh}^2}j_{j_h-1/2}(k_{lh}r)-c_6\sqrt{j_h+1}\frac{(\gamma_1+2\gamma)k_{sh}^2-2\varepsilon}{\gamma k_{sh}^2}j_{j_h-1/2}(k_{sh}r)\}\label{eqS50}
\end{align}
Constants $c_1,c_2,c_3$ in eq. \eqref{eqS45}-\eqref{eqS47} and $c_4,c_5,c_6$ in eq. \eqref{eqS48}-\eqref{eqS50} can be found analytically from BCs (see details in \ref{S3}) and $\varepsilon=m_0E_h/\hbar^2$. Wave numbers of light and heavy holes in $\Gamma_8$ are $k_{lh}$ and $k_{hh}$ respectively. Spin-orbit splited holes in $\Gamma_7$ characterized by $k_{sh}$: 

\begin{eqnarray}\label{eqS51}
k_{lh}^2=\frac{m_0}{\hbar^2(\gamma_1-2\gamma)(\gamma_1+4\gamma)}[2E_h(\gamma_1+\gamma)-\Delta_{\text{SO}}(\gamma_1+2\gamma)+ \nonumber\\ + \{[2E_h(\gamma_1+\gamma)-\Delta_{\text{SO}}(\gamma_1+2\gamma)]^2-\nonumber\\ -4E_h(E_h-\Delta_{\text{SO}})(\gamma_1-2\gamma)(\gamma_1+4\gamma)\}^{\frac{1}{2}}] \, , 
\end{eqnarray}
\begin{eqnarray}\label{eqS52}
k_{sh}^2=\frac{m_0}{\hbar^2(\gamma_1-2\gamma)(\gamma_1+4\gamma)}[2E_h(\gamma_1+\gamma)-\Delta_{\text{SO}}(\gamma_1+2\gamma)- \nonumber\\ - \{[2E_h(\gamma_1+\gamma)-\Delta_{\text{SO}}(\gamma_1+2\gamma)]^2-\nonumber\\-4E_h(E_h-\Delta_{\text{SO}})(\gamma_1-2\gamma)(\gamma_1+4\gamma)\}^{\frac{1}{2}}] \, , 
\end{eqnarray}
\begin{equation}\label{eqS53}
k_{hh}^2=\frac{2m_0E_h}{\hbar^2(\gamma_1-2\gamma)} \, .
\end{equation}
Radial components of even ("+") functions in ZnSe shell have form:
\begin{align}
R^{B,+}_{l_{h_1}}=\mathcal{C}^+_{j_h}\{-[b_{11}j_{j_h+1/2}(\varkappa_{lh}r)+b_{12}y_{j_h+1/2}(\varkappa_{lh}r)]\nonumber\\-[b_{21}j_{j_h+1/2}(\varkappa_{hh}r)+b_{22}j_{j_h+1/2}(\varkappa_{hh}r)]-[b_{31}j_{j_h+1/2}(\varkappa_{sh}r)+b_{32}j_{j_h+1/2}(\varkappa_{sh}r)]\}\label{eqS54}\\R^{B,+}_{l_{h_2}}=\mathcal{C}^+_{j_h}\{-\sqrt{\frac{3(2j_h-1)}{(2j_h+3)}}[b_{11}j_{j_h-3/2}(\varkappa_{lh}r)+b_{12}y_{j_h-3/2}(\varkappa_{lh}r)]\nonumber\\+\sqrt{\frac{(2j_h+3)}{3(2j_h-1)}}[b_{21}j_{j_h-3/2}(\varkappa_{hh}r)+b_{22}y_{j_h-3/2}(\varkappa_{hh}r)]-\sqrt{\frac{3(2j_h-1)}{(2j_h+3)}}[b_{31}j_{j_h-3/2}(\varkappa_{sh}r)+b_{32}y_{j_h-3/2}(\varkappa_{sh}r)]\}\label{eqS55}\\
R^{B,+}_{l_\text{SO}}=\mathcal{C}_{j_h}^+\{c_1\sqrt{\frac{j_h}{(2j_h+3)}}\frac{2(\varepsilon-u)-(\gamma_1+2\gamma)\varkappa_{lh}^2}{\gamma \varkappa_{lh}}[b_{11}j_{j_h+1/2}(\varkappa_{lh}r)+b_{12}y_{j_h+1/2}(\varkappa_{lh}r)]\nonumber\\+\sqrt{\frac{j_h}{(2j_h+3)}}\frac{2(\varepsilon-u)-(\gamma_1+2\gamma)\varkappa_{sh}^2}{\gamma \varkappa_{sh}^2}[b_{31}j_{j_h+1/2}(\varkappa_{sh}r)+b_{31}y_{j_h+1/2}(\varkappa_{sh}r)]\}\label{eqS56}
\end{align}
Radial components of odd ("-") functions in ZnSe shell have form:
\begin{align}
R_{l_{h_1}}^{B,-}=\mathcal{C}^-_{j_h}\{\sqrt{2j_h-1}[b_{41}j_{j_h-1/2}(\varkappa_{lh}r)+b_{42}y_{j_h-1/2}(\varkappa_{lh}r)]+\nonumber\\\sqrt{2j_h-1}[b_{51}j_{j_h-1/2}(\varkappa_{hh}r)+b_{52}y_{j_h-1/2}(\varkappa_{hh}r)]+\sqrt{2j_h-1}[b_{61}j_{j_h-1/2}(\varkappa_{sh}r)+b_{62}y_{j_h-1/2}(\varkappa_{sh}r)]\}\label{eqS57}\\ R_{l_{h_2}}^{B,-}=\mathcal{C}_{j_h}^-\{-\sqrt{3(2j_h+3)}[b_{41}j_{j_h+3/2}(\varkappa_{lh}r)+b_{42}y_{j_h+3/2}(\varkappa_{lh}r)]+\nonumber\\\frac{(2j_h-1)}{\sqrt{3(2j_h+3)}}[b_{51}j_{j_h+3/2}(\varkappa_{hh}r)+b_{52}y_{j_h+3/2}(\varkappa_{hh}r)]-\sqrt{3(2j_h+3)}[b_{61}j_{j_h+3/2}(\varkappa_{sh}r)+b_{62}y_{j_h+3/2}(\varkappa_{sh}r)]\}\label{eqS58}\\ R_{l_{\text{SO}}}^{B,-}=\mathcal{C}_{j_h}^-\{-\sqrt{j_h+1}\frac{(\gamma_1+2\gamma)\varkappa_{lh}^2-2(\varepsilon-u)}{\gamma \varkappa_{lh}^2}[b_{41}j_{j_h-1/2}(\varkappa_{lh}r)+b_{42}y_{j_h-1/2}(\varkappa_{lh}r)]\nonumber-\\\sqrt{j_h+1}\frac{(\gamma_1+2\gamma)\varkappa_{sh}^2-2(\varepsilon-u)}{\gamma \varkappa_{sh}^2}[b_{61}j_{j_h-1/2}(\varkappa_{sh}r)+b_{62}y_{j_h-1/2}(\varkappa_{sh}r)]\}\label{eqS59}
\end{align}
Constants $b_{11},b_{12},b_{21},b_{22},b_{31},b_{32}$ in eq. \eqref{eqS54}-\eqref{eqS56} and $b_{41},b_{42},b_{51},b_{52},b_{61},b_{62}$ in eq. \eqref{eqS57}-\eqref{eqS59} can be found analytically from BCs (see details in \ref{S3}), $y_l(r)$ is Neumann function \cite{Abramowitz1972} and $u= m_0U_B^h/\hbar^2$. Wave numbers of light and heavy holes in $\Gamma_8$ are $\varkappa_{lh}$ and $\varkappa_{hh}$ respectively. Spin-orbit splited holes in $\Gamma_7$ characterized by $\varkappa_{sh}$: 
\begin{eqnarray}\label{eqS60}
\varkappa_{lh}^2=\frac{m_0}{\hbar^2(\gamma_1-2\gamma)(\gamma_1+4\gamma)}[2(E_h-U_{\text{B}}^h)(\gamma_1+\gamma)-\Delta_{\text{SO}}(\gamma_1+2\gamma)+ \nonumber\\ + \{[2(E_h-U_{\text{B}}^h)(\gamma_1+\gamma)-\Delta_{\text{SO}}(\gamma_1+2\gamma)]^2-\nonumber\\ -4(E_h-U_{\text{B}}^h)(E_h-U_{\text{B}}^h-\Delta_{\text{SO}})(\gamma_1-2\gamma)(\gamma_1+4\gamma)\}^{\frac{1}{2}}] \, ,  
\end{eqnarray}
\begin{eqnarray}\label{eqS61}
\varkappa_{sh}^2=\frac{m_0}{\hbar^2(\gamma_1-2\gamma)(\gamma_1+4\gamma)}[2(E_h-U_{\text{B}}^h)(\gamma_1+\gamma)-\Delta_{\text{SO}}(\gamma_1+2\gamma)- \nonumber\\ - \{[2(E_h-U_{\text{B}}^h)(\gamma_1+\gamma)-\Delta_{\text{SO}}(\gamma_1+2\gamma)]^2-\nonumber\\-4(E_h-U_{\text{B}}^h)(E_h-U_{\text{B}}^h-\Delta_{\text{SO}})(\gamma_1-2\gamma)(\gamma_1+4\gamma)\}^{\frac{1}{2}}] \, ,
\end{eqnarray}

\begin{equation}\label{eqS62}
\varkappa_{hh}^2=\frac{2m_0(E_h-U_{\text{B}}^h)}{\hbar^2(\gamma_1-2\gamma)}\, .
\end{equation}
The normalization condition on radial components of hole wave function, which defines constants $\mathcal{C}^{\pm}_{j_h}$ for both even and odd radial wave functions has form:

\begin{equation}\label{eqS63}
\int\limits_0^{+\infty}\left[|R_{l_{h_1}}^{\pm}|^2+|R_{l_{h_2}}^{\pm}|^2+|R_{l_\text{SO}}^{\pm}|^2\right]r^2\text{d}r =1 .
\end{equation}


\newpage

\subsection{Expansion of the Coulomb interaction operator}\label{S6}

To calculate the matrix element of the Coulomb interaction on the exciton wave function, we apply the expansion of the Coulomb potential over the Legendre polynomials:
\begin{eqnarray}\label{eqS64}
\hat V_{\rm Coul}=-\frac{e^2}{\varepsilon_{\rm av}|\bm{r}_e-\bm{r}_h|}=-\frac{e^2}{\varepsilon_{\rm av}}\sum_{l=0}^{\infty}a_l(r_e,r_h)P_l(\cos \vartheta),
\end{eqnarray}
where 
\begin{eqnarray}\label{eqS65}
a_l(r_e,r_h)=r_{<}^{l}/r_{>}^{l+1},\\
P_l(\cos \vartheta)=\frac{4\pi}{2l+1}\sum_{m=-l}^lY_{l,m}^*(\Omega_e)Y_{l,m}(\Omega_h),\label{eqS66}
\end{eqnarray}
and $\vartheta$ is angle between vectors $\bm{r}_e=(r_e,\Omega_e)=(r_e,\theta_e,\varphi_e)$ and $\bm{r}_h=(r_h,\theta_h,\varphi_h)$. For all exciton states, we use only the first term with $l=0$ in the sum \eqref{eqS64}. In the case of excitons formed by the $nS_e$ electron state, only this term gives nonzero contribution to the matrix element of the Coulomb interaction. In the case of excitons formed by the other electron states, this term gives a major contribution to the matrix element of the Coulomb interaction.

The average dielectric constant of an InP/ZnSe NC is defined as $\varepsilon_{\rm av}=(\varepsilon_{\inf}^{\rm InP}+\varepsilon_{\inf}^{\rm ZnSe})/2$, where $\varepsilon_{\inf}^{\rm InP}=9.6$ and $\varepsilon_{\inf}^{\rm ZnSe}=7$ are the high-frequency dielectric constants of InP and ZnSe, respectively.

We need to notice that in reality NC surrounded by organic ligands. This fact can lead to presence of the corrections to the electron and hole quantum size energies as well as to the electron-hole Coulomb interaction. However in the case of spherical-symmetry nanocrystals with infinite barrier at the outer surface these corrections can be calculated perturbatively and compensate each other so that the dielectric confinement effect  doesn't change excitonic transitions energies \cite{Grigoryan1990}.


\subsection{1-PA/2-PA operators}\label{S8}
Shr$\ddot{\text{o}}$odinger equation for electron in the crystal with eigenstates $\mathcal{E}_{n}$ and eigenvectors electron wave functions $\Psi_{n,\bm{k}}$, decomposed by infinite number of Bloch basis functions $u_{n',\bm{k}}$ with $n'$-component envelope wave functions $\varphi^{(n)}$ near dot $\bm{k}=0$ presented below:

\begin{equation}\label{eqS67}
\hat{\mathcal{H}}\Psi_{n,\bm{k}}=\frac{\hbar^2\hat{k}^2}{2m_0}\Psi_{n,\bm{k}}+U_e(\bm{r})\Psi_{n,\bm{k}}=\mathcal{E}_{n}(\bm{k})\Psi_{n,\bm{k}}, \ \Psi_{n,\bm{k}}=\sum_{n'}\varphi^{(n)}_{n'}u_{n',\bm{k}=0}
\end{equation}
In this work the dipole approximation for one- and two-photon absorption analysis is applied. Operator of optical transition can be derived from Schr$\ddot{\text{o}}$dinger equation \eqref{eqS67} with canonical transformation of momentum $\hat{\bm{\pi}}=\hat{\bm{p}}-\frac{e}{c}\bm{A}$ in field $\bm{A}$, i.e. $\hat{\bm{k}}\rightarrow \hat{\bm{k}}-\frac{e}{c\hbar}\bm{A}$ in wave equation \eqref{eqS67}, which leads to equation below:
\begin{eqnarray}\label{eqS68}
\hat{\mathcal{V}}_{\alpha}=-\frac{e}{c}\bm{A}_{\alpha}\cdot\frac{\partial \hat{\mathcal{H}}}{\partial \bm{k}},
\end{eqnarray}
where $\bm{A}_{\alpha}=A_0\bm{e}_{\alpha}$ is the electro-magnetic field vector potential with amplitude $A_0$, which can have circular $\bm{e}_+=1/\sqrt{2}\{1,i,0\}$, $\bm{e}_-=1/\sqrt{2}\{1,-i,0\}$ polarization. However, we can't describe hole and electron simultaneously by one model and infinite number of Bloch basis functions. This leads us to three types of operators. For transition between different Bloch basis functions, for example interband transition between valence state $|v\rangle$ and conduction band $|c\rangle$, the operator \eqref{eqS68} transforms to:
\begin{eqnarray}
\mathcal{V}^{cv}_{\alpha}=-\frac{e}{cm_0}\hat{\bm{p}}\cdot\bm{A}_{\alpha},
\end{eqnarray}
where momentum operator $\hat{\bm{p}}=-i\hbar\bm{\nabla}$ acts on the Bloch components of electron and hole wavefunctions.

As was said previously, we applied $\bm{k}\cdot\bm{p}-$Kane model (see in details (\ref{S2})) with Hamiltonian $\hat{H}_{8\times8}$ to describe electron in conduction band and six-band Luttinger Hamiltonian $\hat{H}_{6\times6}$ to describe holes in valence band. This leads us to two different operators in case of intraband transitions. For conduction band transitions between $|c'\rangle$ and $|c\rangle$ states operator \eqref{eqS68} has form:
\begin{eqnarray}
\hat{\mathcal{V}}^{c'c}_{\alpha}=-\frac{e\hbar}{c}\bm{A}_{\alpha}\cdot \hat{\bm{V}}_{e},
\end{eqnarray}
where $\hat{\bm{V}_e}$ is the same velocity as \eqref{eqS17}. For valence band transitions between $|v'\rangle$ and $|v\rangle$ states operator  takes form:
\begin{eqnarray}
\hat{\mathcal{V}}^{v'v}_{\alpha}=-\frac{e\hbar}{c}\bm{A}_{\alpha}\cdot\hat{\bm{V}}_h,
\end{eqnarray}
where vector operator $\hat{\bm{V}}_h$ components are presented in eq. \eqref{eqS23}-\eqref{eqS25} of SI \ref{S3}.

\subsubsection{Electron and hole states used for calculation of the one- and two-photon optical transitions}\label{S8.1}

\textbf{Electron states:} $nS_e (\times2)$ with $n\in \{1,4\}$, $nP_{e,1/2}(\times2)$ with $n\in \{1,4\}$, $nP_{e,3/2}(\times4)$ with $n\in \{1,4\}$, $nP_{e,1/2}(\times2)$ with $n\in \{1,4\}$,  $nD_{e,3/2}(\times4)$  with $n\in \{1,3\}$, $nD_{e,5/2}(\times6)$ with $n\in \{1,3\}$
\\
\textbf{Hole states:} $nS_{3/2}(\times4)$ with $n\in \{1,8\}$, $nP_{3/2}(\times4)$ with $n\in \{1,8\}$, $nS_{1/2}(\times2)$ with $n\in \{1,6\}$, $nP_{1/2}(\times2)$ with $n\in \{1,6\}$, $nP_{5/2}(\times6)$ with $n\in \{1,6\}$.
\\
Numbers in brackets represent the degeneracy of states. 

\subsection{Influence of material parameters and models on charge carriers energy spectrum}\label{S7}

\begin{figure}[h!]
\begin{center}
\includegraphics[width=1\linewidth]{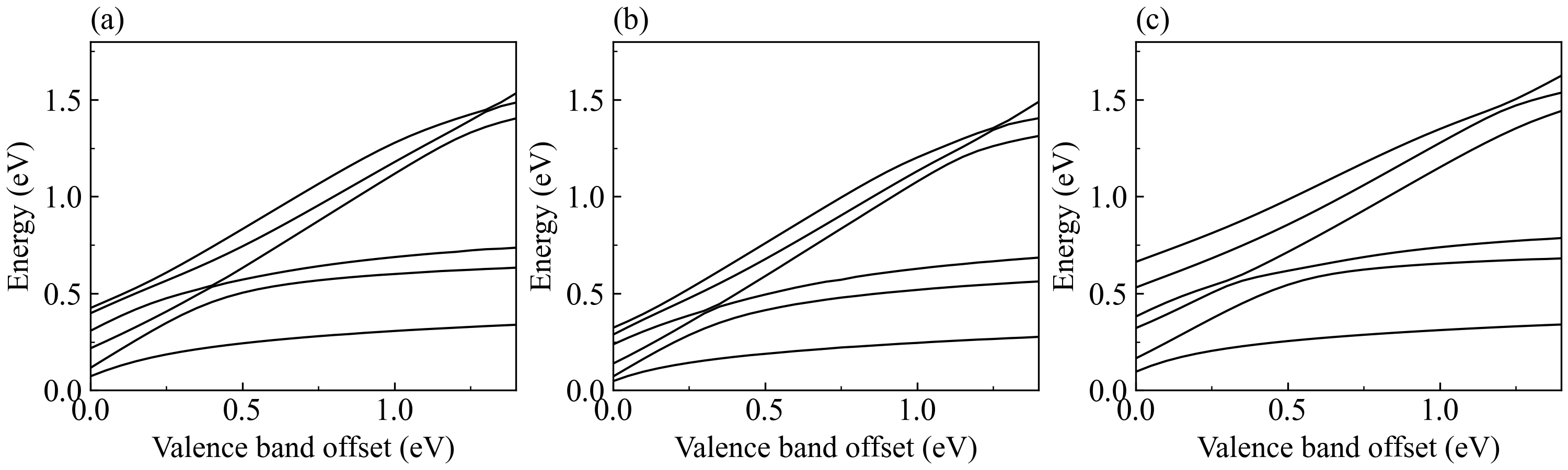}
\caption{Dependence of the energy of the six lowest $S_{3/2}$ hole states on the valence band offset for three sets of the Luttinger parameters $\gamma_1,\gamma_2,\gamma_3$ from (a) Ref.~\cite{Sermage1981} (used for calculations of absorption spectra), (b) Ref.~\cite{Reimann2001} and (c) Ref.~\cite{Venghaus1979}.}
\label{fig:EhCompare}
\end{center}
\end{figure}

The material parameters used for calculations are shown in Table \ref{tab:tableS1}. Dependencies of the energy of the six lowest $S_{3/2}$ states in an InP/ZnSe NC with $a_c=1.45$~nm and $a=4.2$~nm on the valence band offset value for  different sets of Luttinger parameters for ZnSe from Table S1 are shown in Fig.~\ref{fig:EhCompare}. One can see that for all sets of Luttinger parameters, similar energy spectra are obtained, indicating that the choice of Luttinger parameters should not qualitatively modify the obtained absorption spectra. 

\begin{table}[h!]
\centering
\caption{Material parameters used in the calculations: $m_e$ is the electron effective mass in units of the free electron mass $m_0$. $E_g$ is band gap, $\Delta_{\text{SO}}$ is the spin-orbit splitting of the valence band, and $E_p$ is the Kane energy of interaction between valence and conduction bands. $\gamma_1$, $\gamma_2$ and $\gamma_3$ are Luttinger parameters.}
\begin{tabular}{|l|l|l|}
\hline
Parameter                                              & InP      & ZnSe     \\ \hline
$m_e$                                                  & 0.08$m_0$\cite{Vurgaftman2001} & 0.16$m_0$ \cite{Sermage1977} \\
$E_p$                                                  & 20.45 eV\cite{Tadjine2017}   & 24.2 eV \cite{Lawaetz1971}   \\
$E_g$                                                  & 1.423 eV\cite{Vurgaftman2001}   & 2.82 eV \cite{Theis1977}   \\
$\Delta_{\text{SO}}$ & 0.108 eV\cite{Vurgaftman2001}   & 0.403 eV \cite{Sobolev1978}  \\
$\gamma_1$                              & 5.08 \cite{Vurgaftman2001}    & 4.3 \cite{Sermage1981}     \\
$\gamma_2$                              & 1.6\cite{Vurgaftman2001}& 1.1\cite{Sermage1981} \\$\gamma_3$& 2.1\cite{Vurgaftman2001}& 1.84\cite{Sermage1981}  \\
$\gamma_1$                              &      & 2.65\cite{Reimann2001}       \\
$\gamma_2$                              & & 0.63\cite{Reimann2001}   \\$\gamma_3$& & 1.04\cite{Reimann2001}   \\
$\gamma_1$                              &      & 4.8 \cite{Venghaus1979}        \\
$\gamma_2$                              & & 0.67\cite{Venghaus1979} \\$\gamma_3$& & 1.53\cite{Venghaus1979} \\\hline
\end{tabular}\nonumber

\label{tab:tableS1}
\end{table}

\begin{figure}[H]
\begin{center}
\includegraphics[width=0.9\linewidth]{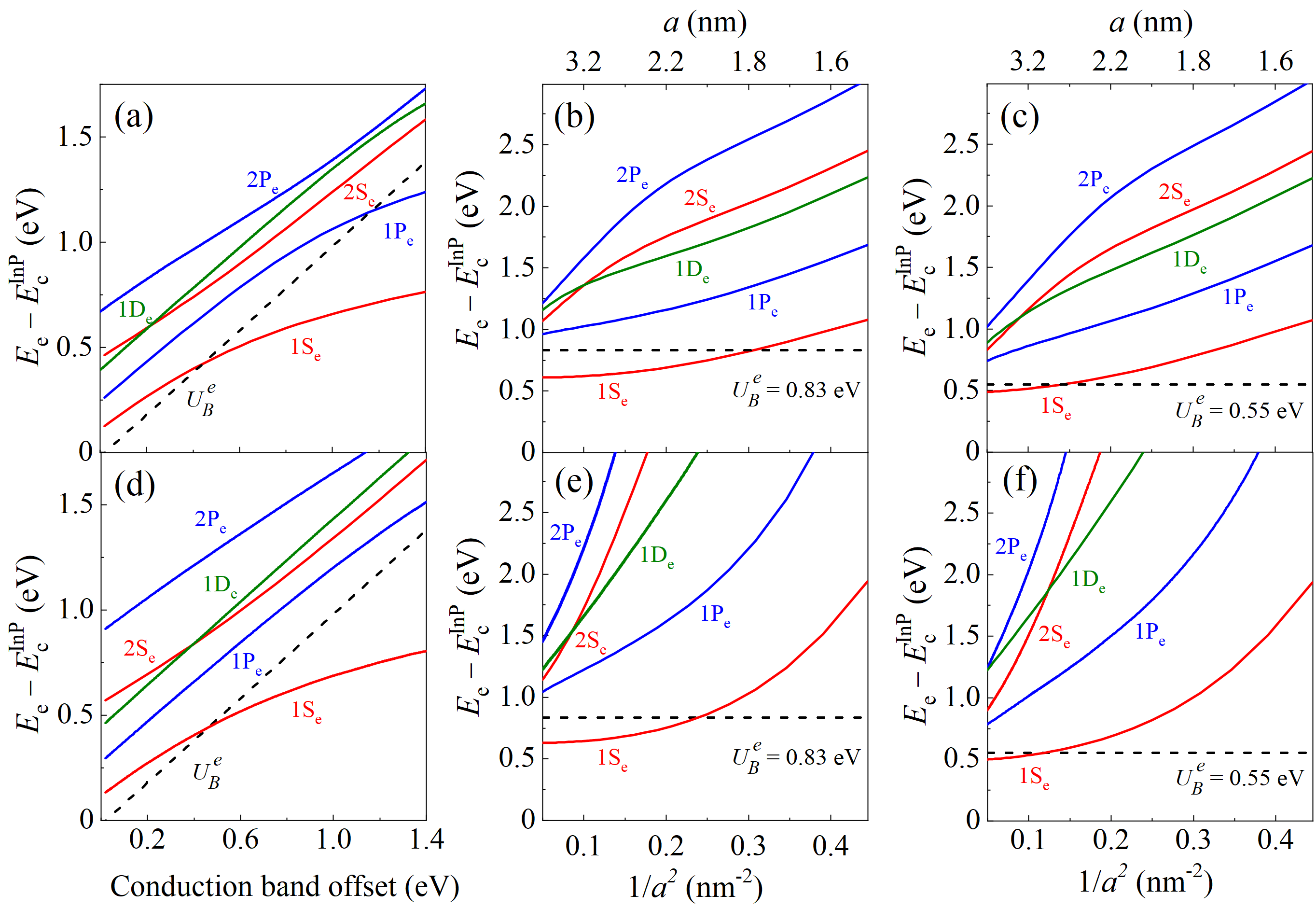}
\caption{Electron energy levels for an InP/ZnSe NC with $a_c=1.45$~nm and $a=4.2$~nm as functions of the band offset $U_B^e$ for (a) the energy-dependent electron mass $m_c(E_e,U_e)$ (b) constant electron mass $m_c(0,0)$. Electron energy levels dependencies on the total NC radius $a$ with fixed core radius $a_c=1.45$ nm for $U_B^e=0.83$~eV are shown in panels (b,e) for the energy-dependent electron mass $m_c(E_e,U_e)$ and constant electron mass $m_c(0,U_e)$, respectively.   Electron energy levels dependencies on the total NC radius $a$ with fixed core radius $a_c=1.45$ nm for $U_B^e=0.55$~eV are shown in panels (c,f) for the energy-dependent electron mass $m_c(E_e,U_e)$ and constant electron mass $m_c(0,0)$, respectively. }
\label{fig:EeCompare}
\end{center}
\end{figure}

Fig  \ref{fig:EeCompare} shows the comparison of the electron energy spectra calculated within the  eight-band Kane model Fig. \ref{fig:EeCompare} (a,b,c) and within the one-band model (d,e,f). The main difference is the overestimation of the excited state energies in the one-band model. One can see that even for NC with $1/a^2\ll 1$, for example, in the case of $a_c=1.45$ nm and $a=4.2$ nm (Fig. \ref{fig:EeCompare} (d)) $1P_e$ level is delocalized for all values of the conduction band offset $U_B^e$, and the energy of the $2P_e$ state is overestimated for about $210$ meV even for $U_B^e=0$ eV. The difference between the two models is especially clear for thin ZnSe shells ($1/a^2\approx0.3-0.4$~$\text{nm}^{-2}$) ( compare Fig. \ref{fig:EeCompare} (b) and (c), Fig. \ref{fig:EeCompare} (e) and (f)).

\newpage

\subsection{One- and two-photon absorption intensities without Coulomb corrections}\label{S9}
\begin{figure}[h!]
\begin{center}
\includegraphics[width=0.8\linewidth]{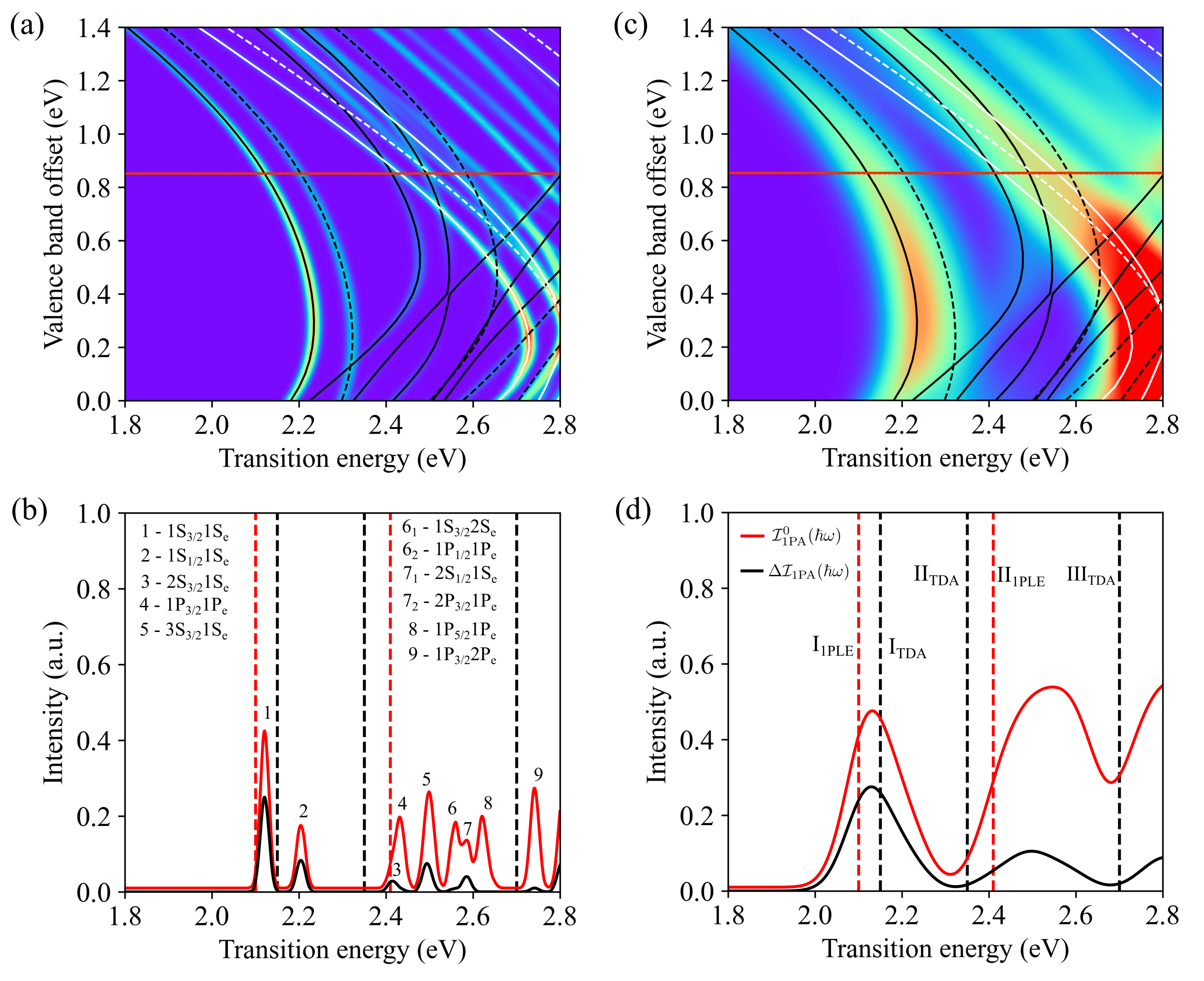}
\caption{ The heat maps of the one-photon absorption intensity ${\cal I}^0_{\rm 1PA}(\hbar\omega)$ without accounting for the Coulomb correction for different valence band offset values calculated for the inhomogeneous broadening of each exciton transition (a) $\Gamma=10$~meV  and (c) $\Gamma=50$~meV, respectively. Red lines in panels (b,d) show slices of the spectra ${\cal I}^0_{\rm 1PA}(\hbar\omega)$ from panels (a,c) for $U_B^h=0.85$~eV (horizontal red line ). Vertical red and black lines ($\rm I_{\rm TDA},\rm II_{\rm TDA}, \rm III_{\rm TDA}$)  correspond to the one-photon transition energies determined in Ref.~\cite{Respekta2025} from 1PLE and differential absorption spectra, respectively. Black lines in panels (b,d) show the calculated differential absorption spectra $\Delta{\cal I}_{\rm 1PA}(\hbar\omega)$. }
\label{fig:Fig1PAnoC}
\end{center}
\end{figure}

In this section, we show one- (Fig.~\ref{fig:Fig1PAnoC}) and two-photon (Fig.~\ref{fig:Fig2PAnoC}) absorption spectra calculated without accounting for the electron-hole Coulomb interaction. 

\begin{figure}[h!]
\begin{center}
\includegraphics[width=0.8\linewidth]{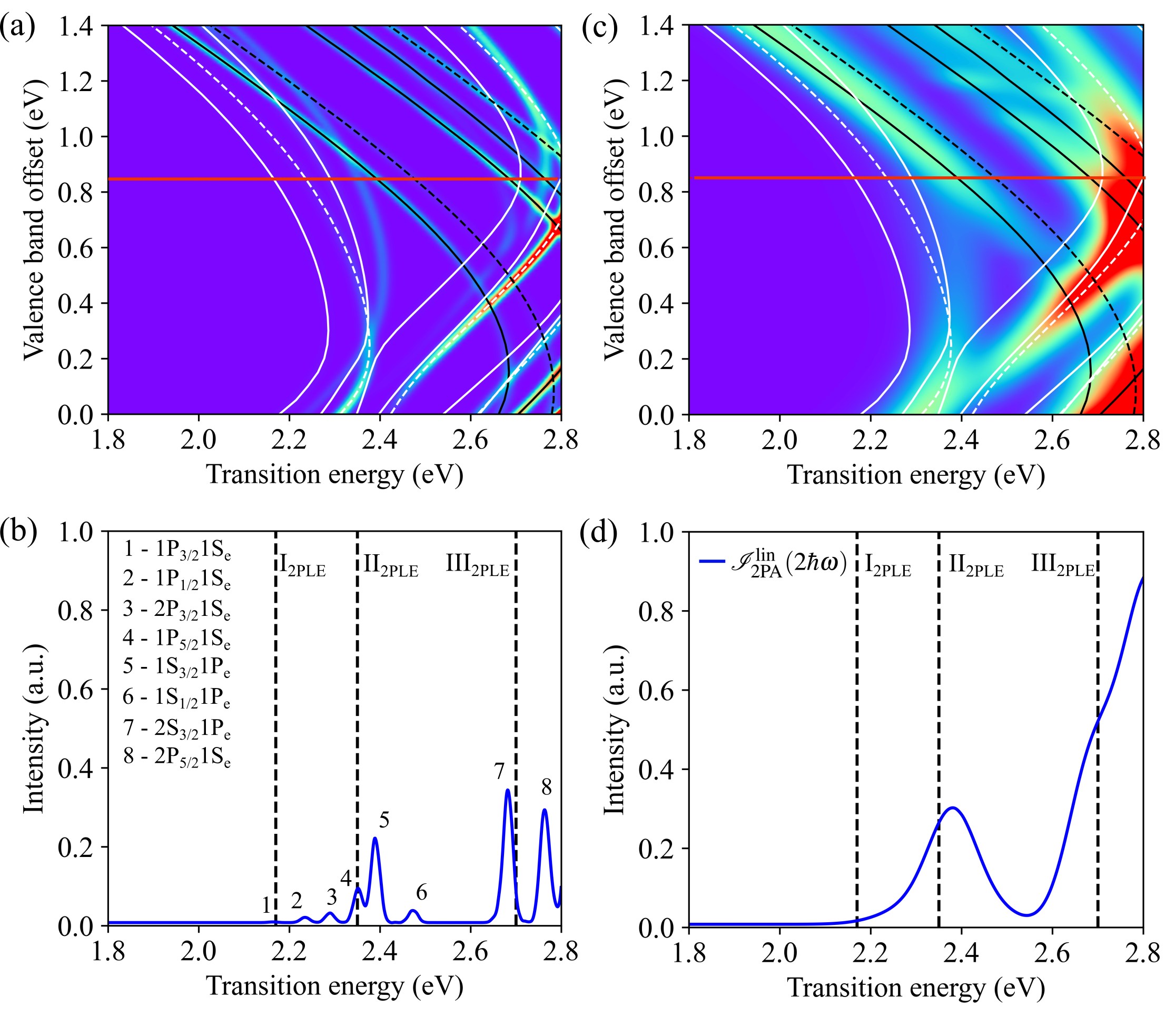}
\caption{  The heat map of the linearly-polarized two-photon absorption intensity ${\cal I}_{\rm 2PA}^{\rm lin}(2\hbar\omega)$ calculated without Coulomb corrections for different valence band offset values, calculated for the inhomogeneous broadening of each exciton transition  (a) $\Gamma=10$~meV and (c) $\Gamma=50$~meV, respectively. Solid and dashed white lines correspond to $nP_{3/2}1S_e$ and $nP_{1/2}1S_e$ exciton energies, respectively. Solid and dashed black lines correspond to $nS_{3/2}1P_e$ and $nS_{1/2}1P_e$ exciton energies, respectively. Panels (c,d) show cutoffs of the heat maps from panels (a,b) at the level $U_B^h=0.85$~eV (horizontal red line). Vertical dashed lines correspond to the two-photon transition energies ($\rm I_{\rm 2PLE},\rm II_{\rm 2PLE},\rm III_{\rm 2PLE}$) observed in Ref.~\cite{Respekta2025}.  }
\label{fig:Fig2PAnoC}
\end{center}
\end{figure}

\begin{figure}[h!]
\begin{center}
\includegraphics[width=1\linewidth]{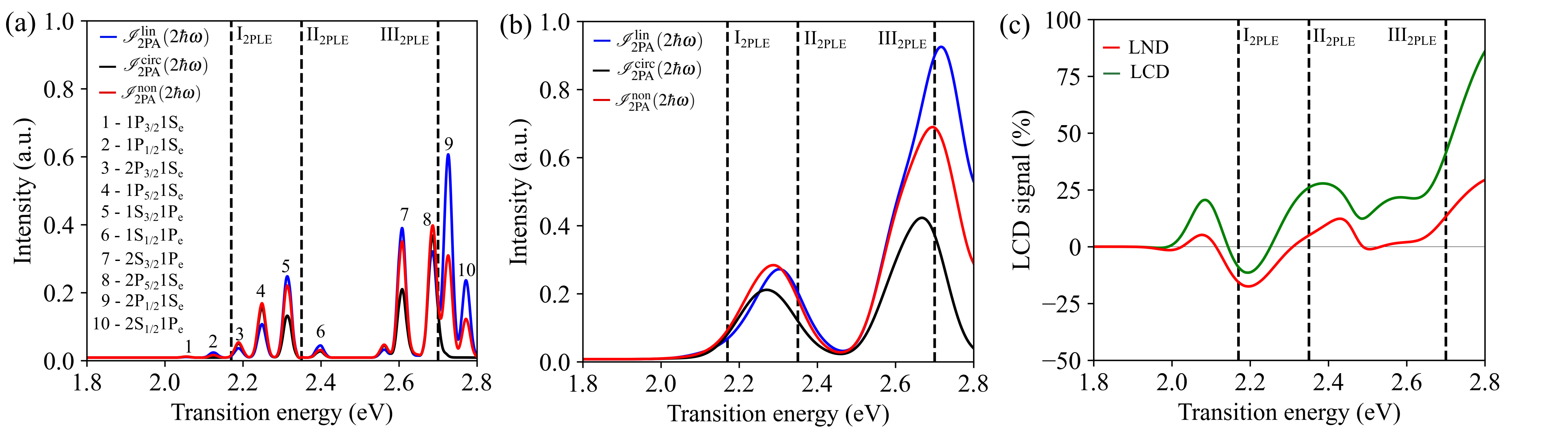}
\caption{   The two-photon absorption spectrum in the case of linear ${\cal I}_{\rm 2PA}^{\rm lin}(2\hbar\omega)$ (blue) and circular  ${\cal I}_{\rm 2PA}^{\rm circ} (2\hbar\omega)$ (black), polarization of photons, calculated for the inhomogeneous broadening of each exciton transition (a) $\Gamma=10$~meV and (b) $\Gamma=50$~meV, respectively.  Nonpolarized absorption spectra ${\cal I}_{\rm 2PA}^{\rm non} (2\hbar\omega)$ are shown by red lines. (c) Spectral dependence of the linear-circular dichroism signal (LCD) and the linear-nonpolarized dichroism signal (LND) for absorption spectra from panel (b). }
\label{fig:FigLCDSI}
\end{center}
\end{figure}
   
\subsection{Influence of the lattice strain on the one-photon absorption spectra}\label{S10}
\begin{figure}[H]
\begin{center}
\includegraphics[width=0.8\linewidth]{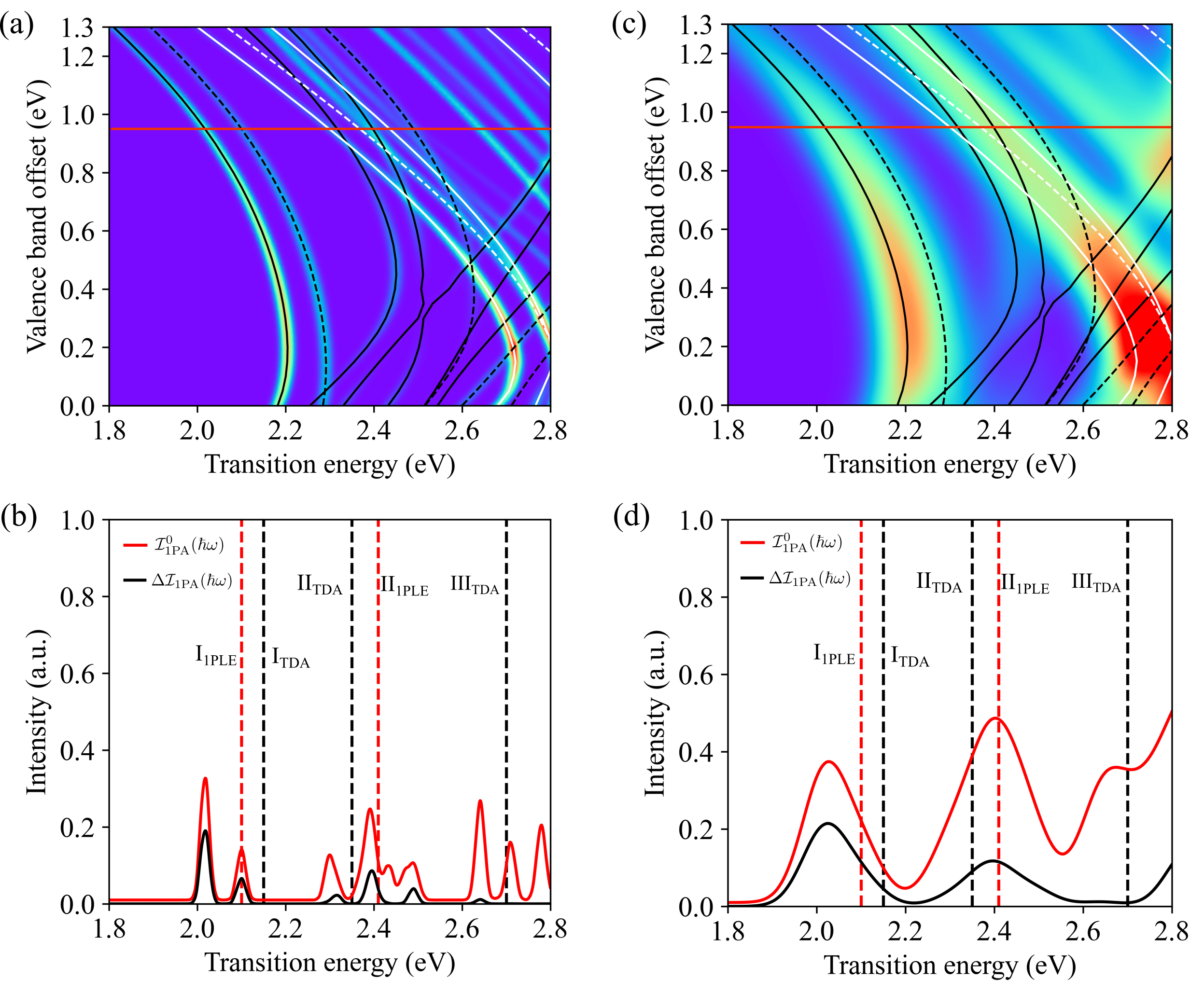}
\caption{ The heat maps of the one-photon absorption intensity ${\cal I}^0_{\rm 1PA}(\hbar\omega)$for different valence band offset values calculated for the inhomogeneous broadening of each exciton transition (a) $\Gamma=10$~meV  and (c) $\Gamma=50$~meV, respectively. Red lines in panels (b,d) show slices of the spectra ${\cal I}^0_{\rm 1PA}(\hbar\omega)$ from panels (a,c) for $U_B^h=0.95$~eV (horizontal red line ). Vertical red and black lines ($\rm I_{\rm TDA},\rm II_{\rm TDA}, \rm III_{\rm TDA}$)  correspond to the one-photon transition energies determined in Ref.~\cite{Respekta2025} from 1PLE and differential absorption spectra, respectively. Black lines in panels (b,d) show the calculated differential absorption spectra $\Delta{\cal I}_{\rm 1PA}(\hbar\omega)$. The lattice strain resulting in $0.1$~eV increase of the InP band gap is included. }
\label{fig:Fig2PAnoC}
\end{center}
\end{figure}

\end{document}